\documentclass{jfm}

\usepackage{graphicx}
\usepackage{newtxtext}
\usepackage{newtxmath}
\usepackage{bm} 
\usepackage{natbib}
\usepackage{hyperref}
\hypersetup{
    colorlinks = true,
    urlcolor   = blue,
    citecolor  = black,
}

\newcommand{\RomanNumeralCaps}[1]
\linenumbers

\title{Statistics of velocity gradient and vortex sheet structures in polymeric turbulent von K{\'a}rm{\'a}n swirling flow}

\author{Feng Wang\aff{1},
	Yi-Bao Zhang \aff{1,2},
	Ping-Fan Yang \aff{1}\corresp{\email{yangpingfan@nwpu.edu.cn}}
	\and Heng-Dong Xi\aff{1}\corresp{\email{hengdongxi@nwpu.edu.cn}}}

\affiliation{\aff{1}Institute of Extreme Mechanics, School of Aeronautics, National Key Laboratory of Aircraft Configuration Design, Key Laboratory for Extreme Mechanics of Aircraft of Ministry of Industry and Information Technology, Northwestern Polytechnical University, Xi’an 710072, Shaanxi, China
\aff{2}New Cornerstone Science Laboratory, Center for Combustion Energy, Key Laboratory for Thermal Science and Power Engineering of Ministry of Education, Department of Energy and Power Engineering, Tsinghua University, Beijing 100084, China}

\begin{document}
\maketitle

\begin{abstract}
Investigations into the effects of polymers on small-scale statistics and flow patterns were conducted in a turbulent von K{\'a}rm{\'a}n swirling (VKS) flow. We employed the tomographic particle image velocimetry (Tomo-PIV)  technique to obtain full information on three-dimensional velocity data, allowing us to effectively resolve dissipation scales. Under varying Reynolds numbers ($R_\lambda=168\text{-}235$) and polymer concentrations ($\phi=0\text{-}25~\rm ppm$), we measured the velocity gradient tensor (VGT) and related quantities. Our findings reveal that the ensemble average and probability density function (PDF) of VGT invariants, which represent turbulent dissipation and enstrophy along with their generation terms, are suppressed as polymer concentration increases. Notably, the joint PDFs of the invariants of VGT, which characterize local flow patterns, exhibited significant changes. Specifically, the third-order invariants, especially the local vortex stretching, are greatly suppressed, and strong events of dissipation and enstrophy coexist in space. The local flow pattern tends to be two-dimensional, where the eigenvalues of the rate-of-strain tensor satisfy a ratio $1:0:-1$, and the vorticity aligns with the intermediate eigenvector of the rate-of-strain tensor while is perpendicular to the other two. We find that these statistics observations can be well described by the vortex sheet model. Moreover, we find that these vortex sheet structures align with the symmetry axis of the VKS system and orient randomly in the horizontal plane. Further investigation, including flow visualization and conditional statistics on vorticity, confirms the presence of vortex sheet structures in turbulent flows with polymer additions. Our results establish a link between single-point statistics and small-scale flow topology, shedding light on the previously overlooked small-scale structures in polymeric turbulence.

\end{abstract}



\section{Introduction}
Polymers have long been recognized as highly effective drag-reducing agents, a discovery first reported by \cite{toms1948some}. Despite their significant impact on chemical engineering and marine applications, the underlying mechanism behind polymer-induced turbulent drag reduction remains elusive. In the 1980s, de Gennes proposed the elastic theory, suggesting that stretched polymers can exchange energy with turbulent flows \citep{de1986towards}. This exchange disrupts the energy transfer balance across scales \citep{frisch1995turbulence}. Over the past four decades, numerous studies have explored how polymers affect the energy cascade in bulk turbulence, including theoretical \citep{sreenivasan2000onset, fouxon2003spectra, casciola2007energy}, experimental \citep{ouellette2009bulk, xi2013elastic, de2016concentration, zhang2021experimental, zhang2022measured, wang2025energy} and numerical works \citep{benzi2003shell, de2005homogeneous, kalelkar2005drag, perlekar2006manifestations, perlekar2010direct, cai2010dns, watanabe2013hybrid, watanabe2014power, valente2014effect,valente2016energy,nguyen2016small, fathali2019spectral,ur2022effect,rosti2023large}. 
The violent energy extraction by polymers is reflected in the suppression of the velocity structure functions and emergence of the elastic scaling range which is between dissipation range and the inertial range as displayed in the recent experimental study \citep{zhang2021experimental}, and was subsequently reproduced numerically \citep{rosti2023large}. Despite this progress, there remains an open question regarding how polymers specifically impact turbulent dissipation scales. Notably, there is a scarcity of research exclusively focused on small-scale properties in polymeric turbulence, which constitutes the overarching goal of this paper.

The small-scale properties of turbulent flows can be effectively understood through the velocity gradient tensor (VGT), denoted as $A_{ij}\equiv \partial u_i /\partial x_j$ \citep{meneveau2011lagrangian}, where $\mathbf{u}$ represents the velocity field. The VGT provides valuable insights into fine-scale flow motions. For instance, in incompressible turbulence, the dissipation and production of small scales can be expressed in terms of the second- and third-order moments of $\mathbf{A}$, namely its second- and third-order invariants $Q \equiv -\frac{1}{2} A_{ij}A_{ji}$ and $R\equiv -\frac{1}{3} A_{ij}A_{jk}A_{ki}$. As a second-rank tensor, $\mathbf{A}$ can be naturally decomposed into a symmetric part $S_{ij} \equiv \frac{1}{2}\left(A_{ij} + A_{ji} \right) $ and an antisymmetric part $\Omega_{ij} \equiv \frac{1}{2}\left(A_{ij} - A_{ji} \right) $. $\mathbf{S}$ and $\mathbf{\Omega}$ are also known as the rate-of-strain tensor and the rate-of-rotation tensor, which describe the deformation and rotation motion in turbulent flows, respectively. The latter is related to the vorticity vector $\bm{\omega}=\nabla\times\mathbf{u}$ through $\Omega_{ij}=-\frac{1}{2} \epsilon_{ijk} \omega_k$, where $\epsilon_{ijk}$ is the Levi-Civita symbol. To visualize coherent structures, such as worm-like vortex structures and surrounding dissipation features in Newtonian turbulence, it is helpful to consider the norm of $\mathbf{S}$ and $\mathbf{\Omega}$, that is, $S_{ij} S_{ij}$, which is just the dissipation by multiplying the viscosity, and the enstrophy $\omega_i \omega_i$ \citep{kerr1985higher, she1990intermittent, ganapathisubramani2008investigation, buaria2022vorticity}. The VGT and related quantities thus have gained popularity for analyzing small-scale structures in various turbulent flows, such as wall-bounded flows \citep{blackburn1996topology, chong1998turbulence}, isotropic bulk flows \citep{jimenez1993structure, carter2018small}, jet flows \citep{da2008invariants, ganapathisubramani2008investigation}, wake flows \citep{gomes2014evolution, buxton2017invariants}, Rayleigh-Bénard convection flow \citep{xu2024experimental}, etc.. Consistent with the K41 theory \citep{kolmogorov1941local}, the small-scale statistics of different types of Newtonian turbulent flows exhibit similar features. In this context, we highlight two relevant features closely related to our work. First, there is a preferred alignment of the vorticity vector with the intermediate eigenvector of the rate-of-strain tensor. Second, the joint probability density distributions of the invariants $Q $ and $R$, also known as the PDF on the $R$-$Q$ map, assume a teardrop shape. These inherent features, as summarized in \cite{meneveau2011lagrangian}, are associated with the rich phenomena of small-scale motions that effectively distinguish turbulent flows from other random velocity fields.

According to the discussions above, it’s evident that the impact of polymers on the small scales of turbulence can be assessed by analyzing the statistics of the VGT. First of all, the dissipation $S_{ij} S_{ij}$ and enstrophy $\omega_i \omega_i$, representing the magnitude of VGT, are reduced by polymers and in particular, the reduction in dissipation can be regarded as the drag reduction in bulk turbulence \citep{kalelkar2005drag, perlekar2006manifestations, cai2010dns, perlekar2010direct}. Moreover, the probability distribution of these quantities also contracts after adding polymers, following the same trend as ensemble-averaged values \citep{liberzon2005turbulence, liberzon2006turbulent, cai2010dns, perlekar2010direct, watanabe2013hybrid, cocconi2017small, zhang2022measured}. On the other hand, these distributions for different polymer concentrations collapse when normalized by their standard deviations, suggesting that their functional forms are independent of polymer additives \citep{watanabe2013hybrid, zhang2022measured}. Additionally, by examining the effect of polymers on the $R$-$Q$ map, Liberzon found that the typical teardrop shape remains mostly unchanged, except for a reduction in the tail area \citep{liberzon2006turbulent}. Direct numerical simulation (DNS) study by \cite{perlekar2010direct} reported similar observations as the experimental results in \cite{liberzon2006turbulent}. Beyond single-point statistics, polymers also influence the local flow pattern, and the most striking observation is that the vortex filaments are significantly suppressed, leading to much smoother streamlines with weaker velocity gradients \citep{de2016concentration, zhang2022measured, rosti2023large}. More precisely, this suppression manifests as a reduction in coherent structures, such as vortex tubes or filaments, particularly when viscoelasticity is enhanced \citep{bonn1993small, perlekar2006manifestations, perlekar2010direct, cai2010dns, ur2022effect}. Besides, it is also found that additional viscoelasticity can lead to vortex tube structures with increased dimensions \citep{de2005homogeneous}, primarily in width compared to length \citep{watanabe2013hybrid}. 

Most studies on small-scale processes in polymeric turbulence rely on numerical simulations, while experimental investigations are relatively rare due to technological constraints. For instance, \cite{bonn1993small} utilized the visibility of bubbles to reveal the suppression of vortex filaments in polymeric turbulence, but no information about the VGT was obtained. Using the particle tracking velocimetry (PTV) method, \cite{liberzon2005turbulence, liberzon2006turbulent} measured the statistics of VGT, but the data points were sparse in space due to the low concentration of tracer particles, which restricted the observations of Eulerian flow structures. In many cases, planar particle image velocimetry (PIV) is commonly employed \citep{cocconi2017small, zhang2022measured}, but it provides only two-dimensional velocity information in a plane, which is insufficient for calculating all components of the VGT and for characterizing flow structures. To address this gap, our prior work \citep{wang2025tomo} utilized tomographic PIV (Tomo-PIV) to obtain all components of VGT and enable the visualization of three-dimensional flow structures. Remarkably, most of the experimental and numerical discussions mentioned above are restricted by the relatively low Reynolds numbers. Specifically, the Reynolds number based on the Taylor microscale typically does not exceed $100$. These limitations arise mainly due to finite spatial resolution or computational constraints. In our study, we employ the Tomo-PIV method and provide a dataset with a relatively high Reynolds number, which could further investigate the role of polymers.

In this paper, we study the statistics of VGT in polymeric turbulence in detail. We measure the VGT by Tomo-PIV method in the bulk of a turbulent von K{\'a}rm{\'a}n swirling (VKS) flow. Our Tomo-PIV datasets allow us to reconstruct the three-dimensional Eulerian velocity field with sufficient spatial resolution for calculating all components of the velocity gradients. Focusing on scales dominated by energy dissipation, this study enriches our understanding of how polymers impact the generation of small-scale structures. We present the experimental details, including setup and data acquisition, in section \ref{sec:exp}. Section \ref{sec:stat} focuses on VGT-based small-scale statistics and explores their dependence on polymer concentration and turbulent Reynolds number. We will see that in addition to the attenuation of magnitudes, we can observe significant changes in small-scale flow topology, hinting at the presence of vortex sheet structures in polymeric turbulence. To further validate our observations, we provide instantaneous flow visualizations in section \ref{sec:visual}. Next, in section \ref{sec:cond}, we present conditional statistics on high enstrophy, which reinforce our findings in the previous sections. Finally, we conclude in section \ref{sec:conclusion} with a summary of our study.

\section{Experimental details} \label{sec:exp}
The experimental arrangement is the same as our prior work \citep{wang2025tomo}, and here we provide a short but crucial introduction, including the setup, measurement method, data collection, and the definition of parameters. For further details, readers can refer to our previous work \citep{zhang2021experimental, wang2025tomo, wang2025energy}.

The VKS flow system serves as our turbulence generator, which was dedicated to the study of polymer-laden turbulence with high Reynolds number \citep{bonn1993small, cadot1998turbulent, crawford2008fluid, ouellette2009bulk, xi2013elastic, burnishev2016influence, de2016concentration, sinhuber2018probing, zhang2021experimental, zhang2022measured}. The VKS flow is driven by two counter-rotating baffled disks located face-to-face on the base plates at both ends of the closed cylindrical container. The container is 636~mm in height and 480~mm in diameter, with a transparent plexiglass side wall that is 10 mm thick. Seven flat optical windows are mounted on the side wall to facilitate optical measurement, and one of these windows is removable for access inside the container. The disks themselves have a radius of 110 mm and are equipped with eight vanes, each 50 mm in height, to enhance mixing. These two disks are 416 mm apart,  along with the surrounding sidewalls, creating an area with an aspect ratio of approximately 1 at the center of the VKS setup. Each disk is driven by a 1.5 kW servo motor, allowing for continuous variation of the rotation frequency, denoted as $f$. The working fluid’s temperature is precisely controlled at $20^\circ$C with minor variation (less than $0.1^\circ$C) using a refrigerated circulator and monitored via a PT100 probe inside the container. To achieve the necessary resolution for precise calculation of VGT, we increased the fluid viscosity by using glycerol. In this study, the Newtonian case denotes the aqueous solution of glycerol with a mass fraction of 35\%, resulting in a viscosity of $\nu=2.75\times10^{-6} ~ \rm m^2 s^{-1}$. We varied the disk rotation frequency across three cases: $f=0.45,0.65,0.85~\rm Hz$, allowing us to investigate the Reynolds number ($Re$) dependence, where $Re=(2\pi rf)r/\nu$.

In the polymeric case, we utilize polyacrylamide (PAM) polymer with an average molecular weight of $M_w=18\times10^6$ atomic mass unit (a.m.u.).  To prepare the concentrated stock solution of PAM, we followed a similar method reported in our previous study \citep{zhang2021experimental}. There is an essential difference: instead of using deionized water, we employed the aforementioned aqueous solution of glycerol as the solvent. Subsequently, we slowly added the stock solution containing polymers of a specific quality into the VKS via gravity. Light stirring was carried out before the formal experiments to ensure that the polymer solutions were well mixed yet minimized the physical degradation of PAM. The relaxation time $\tau_p$, which characterizes the viscoelasticity of flexible polymer, can be estimated as 115~ms using the Zimm model \citep{crawford2004particle}. The polymer concentration $\phi$ ranges from 2 to 25 parts per million (ppm) by weight, which falls within the dilute solution regime, where the effects on fluid viscosity can be safely neglected. Also, notice that for polymeric cases with different $\phi$, we still use the parameters of their Newtonian counterparts, such as the Reynolds number, to discriminate these flow fields.

 Our objective in this study is to investigate the impact of polymers on bulk turbulence. Consequently, we focus mainly on the flow at the center of the VKS system, which remains unaffected by side walls. The observation window for the velocity field is measured using Tomo-PIV method and has a size of $31\times29\times6 ~\rm mm^3$ in $\bm{e}_x,\bm{e}_y,\bm{e}_z$ direction, respectively, where $\bm{e}_y$ denotes the vertical direction (which is the axisymmetric axis of VKS system) and $\bm{e}_x$ and $\bm{e}_z$ lies in the horizontal plane. Four cameras, two in the horizontal (or equatorial) plane (the $\bm{e}_x O\bm{e}_z$ plane) and two in the vertical (or meridian) plane (the $\bm{e}_yO \bm{e}_z$ plane), were arranged at an angle of 30 degrees around the $\bm{e}_z$ axis, forming a cross-like configuration \citep{scarano2012tomographic}. Each camera had a 100 mm focal length lens and a 2X extender mounted on the Scheimpflug adapter. The seeding particles used are hollow glass spheres with a density of $\rho_0=1.1\times10^3 ~\rm kg/m^3$ and a mean diameter of $d_0=10 ~\rm \mu m$. These inertia-less particles are illuminated by a Nd:YLF laser with a wavelength of 527~nm. A mirror was used to enhance the intensity and uniformity of the laser beam. Light passing through the optical windows was simultaneously imaged and collected by all four cameras. Calibration involved a 2-level target plate and subsequent refinement through volume self-calibration  \citep{wieneke2008volume}, achieving a disparity lower than 0.1 pixel. The collected images undergo several processing steps, including image pre-processing, volume reconstruction (using the MART technique), volume cross-correlation, and vector post-processing. The final interrogation volumes consisted of 32 voxels on each side, overlapping at 50\%. The vector spacing $dx$ is thus $dx=0.55~\rm mm$, with $57\times51\times11$ velocity vectors in $\bm{e}_x$, $\bm{e}_y$, and $\bm{e}_z$ directions, respectively. For each case with different $Re$ and $\phi$, we collected 3000 snapshots to achieve statistical convergence, with number of statistics approximately $10^8$. All Tomo-PIV procedures mentioned above are executed using DaVis 8.4.0 software.

From the measured velocity field, we have examined the mean flow, which consists of a pumping motion and a shear motion, as reported in previous VKS experiments \citep{voth2002measurement, debue2018experimental}. Unless otherwise specified, we adopt the fluctuating velocity $u_i$ from a classical Reynolds decomposition for data analysis and refer to it simply as velocity, where $i=x,y,z$ corresponds to laboratory coordinates. In addition, $\mathbf{u}$ exhibits a larger amplitude and better homogeneity than the mean flow, allowing us to use the spatial average to replace the ensemble average. Before doing data analysis, it is crucial to test the volumetric measurement's accuracy quantitatively. The three velocity components are independently determined, each accompanied by random errors. Therefore, evaluating how well the datasets satisfy the divergence-free condition $\partial u_i/\partial x_i=0$ is a valuable test for the accuracy of the measurement \citep{ganapathisubramani2007determination, lawson2014scanning, fiscaletti2022tomographic}. The correlation coefficient between $-(\partial u_z/\partial z)$ and $\partial u_x/\partial x+\partial u_y/\partial y$ is calculated to be around 0.8 in all cases and increases by about $10\%$ after removing the boundary data points which have more significant errors, the details of this method are discussed in our previous work \citep{wang2025tomo}. Consequently, we can directly calculate VGT and related quantities from the measured $\mathbf{u}$ fields, enabling analysis of small-scale information of turbulence with and without polymers.

After conducting thorough qualitative and quantitative assessments, we obtained datasets that allowed us to perform statistical and structural analyses of small-scale turbulence.  For these data sets, we use the root-mean-square (rms) of the fluctuating velocity to measure the turbulent characteristic velocity, defined as $u^\prime=  \left\langle \frac{1}{3} u_i (\mathbf{x},t) u_i (\mathbf{x},t)\right\rangle^{1/2} $. Since the flow field is statistically stationary and homogeneous, we employ the temporal and spatial averages in place of the ensemble average, denoted by the angled bracket $\left\langle~\right\rangle$ in the following text. 
In addition to the aforementioned Reynolds number $Re$ based on the geometry of the VKS system, we also use the Reynolds number based on the Taylor microscale, defined as $R_\lambda = \left( 15u^{\prime 4}/\varepsilon \nu \right)^{1/2}$, and these two Reynolds number satisfy the relation $R_\lambda \propto \sqrt{Re}$ \citep{pope2000turbulent}. Here,
$\varepsilon =2\nu \left\langle S_{ij} S_{ij} \right\rangle $ denotes the mean energy dissipation rate and along with the viscosity $\nu$ characterize the dissipation properties of turbulent flow \citep{kolmogorov1941local}. 
The Kolmogorov length and time scales are thus defined as $\eta=\left( \nu^3/\varepsilon \right)^{1/4}$ and $ \tau_\eta = \left( \nu/\varepsilon \right)^{1/2}$, respectively. For the polymeric case, an additional parameter comes into play: the Weissenberg number $Wi=\tau_p / \tau_\eta$, which characterizes the relative strength between elastic force and inertial force. Since $\tau_p$ remains constant while $\tau_\eta$ varies with $R_\lambda$, $Wi$ is coupled with $R_\lambda$. In table \ref{tab:para}, we summarize and list all the parameters mentioned above for our experimental datasets. For all experimental datasets, the values of $Wi$ are larger than one and satisfy Lumley’s time criterion \citep{lumley1969drag}, which suggests that the polymer can be stretched by the local flow when the flow responds faster than the polymer. The relative spatial resolution $dx/\eta$ has an order of magnitude of one and is also included in the table.

\begin{table}
	\begin{center}
		\def~{\hphantom{0}}
		\begin{tabular}{lccc}
			$f~\rm (Hz)$ 								& 0.45 	& 0.65 	& 0.85 \\
			Reynolds number $Re~(\times10^4)$		    & 1.24	& 1.80	& 2.35 \\
			Taylor Reynolds number $R_\lambda$  		& 168	& 203	& 235  \\
			Kolmogorov length scale $\eta~\rm (mm)$ 	& 0.36  & 0.27	& 0.22 \\
			Kolmogorov time scale $\tau_\eta~\rm (ms)$  & 46.4 	& 26.4	& 18.1 \\
			Weissenberg number $ Wi $ 					& 2.5	& 4.3 	& 6.3 \\
			$dx/\eta$ 									& 1.59	& 2.07	& 2.52 \\
		\end{tabular}
		\caption{Parameters of the experimental datasets in this study.}
		\label{tab:para}
	\end{center}
\end{table}

\section{Statistics of velocity gradient tensor} \label{sec:stat}
In this section, we focus on the statistical properties of VGT $\mathbf{A}$, including its invariants, the alignments between vorticity $\bm{\omega}$ and the eigendirections of $\mathbf{S}$, and their orientations in the laboratory frame, etc.. We will see that our experimental results indicate the existence of vortex sheet structures in polymeric turbulence, which is distinct from its Newtonian counterpart. To begin, We first briefly introduce some definitions and properties of VGT $\mathbf{A}$.

The VGT $\mathbf{A}$ has three invariants, $P$, $Q$ and $R$. This work only considers incompressible flows, where the first invariant $P=-A_{ii}$ becomes zero due to the incompressible constraint $\partial u_i / \partial x_i = 0$. Consequently, the remaining two invariants, $Q \equiv -\frac{1}{2} A_{ij}A_{ji}$ and $R\equiv -\frac{1}{3} A_{ij}A_{jk}A_{ki}$, characterize the structure of VGT, and each point on the $R$-$Q$ map represents a specific local flow pattern. Similarly, $\mathbf{S}$ and $\mathbf{\Omega}$ have their corresponding invariants, of which the non-zero ones are $Q_s\equiv -\frac{1}{2} S_{ij} S_{ij}$ and $R_s \equiv -\frac{1}{3} S_{ij} S_{jk} S_{ki}$ for $\mathbf{S}$, and $Q_\omega \equiv \frac{1}{4} \omega_i \omega_i$ for $\mathbf{\Omega}$. Notice that the dissipation $S_{ij} S_{ij}$ and enstrophy $\omega_i \omega_i$ mentioned before are actually equivalent to the second-order invariants $Q_s$ and $Q_\omega$, respectively, and the third-order invariants, the strain-self amplification $R_s$ and the vortex stretching $R_\omega \equiv -\frac{1}{4} \omega_i S_{ij} \omega_j$, appear as the nonlinear terms in the equations of  $Q_s$ and $Q_\omega$ \citep{meneveau2011lagrangian,johnson2024multiscale}. Now noticing that $\mathbf{A}\equiv \mathbf{S}+\mathbf{\Omega}$, and the quantities mentioned above satisfy the following relations:

\begin{equation} \label{eq:Q_def}
	Q = \frac{1}{4} \left( \omega_i \omega_i - 2S_{ij} S_{ij} \right) = Q_\omega + Q_s
\end{equation}
\begin{equation} \label{eq:R_def}
	R = -\frac{1}{3} \left( S_{ij} S_{jk} S_{ki} + \frac{3}{4}\omega_i S_{ij} \omega_j \right) = R_s + R_\omega.
\end{equation}

Hence we can see that $Q$ represents the relative strength between dissipation and enstrophy, while $R$ represents the relative strength between strain self-amplification and vortex stretching. These invariants can characterize specific local flow topologies \citep{soria1994study, blackburn1996topology,meneveau2011lagrangian,johnson2024multiscale}. For instance, the $R$-$Q$ map is a typical example, and other maps, such as those between $R_s$ and $Q_s$ and between $Q_s$ and $Q_\omega$, serve as other examples. We will discuss them in detail in the following text. In particular, the eigenvalues of $\mathbf{S}$ are defined as $\lambda_i$  and the corresponding eigenvectors as $\bm{e}_i$, where $i=1,2,3$, and the eigenvalues are ordered as $\lambda_1>\lambda_2>\lambda_3$. Due to the incompressible condition, we have $\lambda_1+\lambda_2+\lambda_3=0$, thus the extensive eigenvalue $\lambda_1\geq 0$ while the compressive one $\lambda_3 \leq 0$, and the intermediate one $\lambda_2$ can be either positive or negative. The vortex stretching $\omega_i S_{ij}\omega_j$ can be well understood through the alignment of $\bm{\omega}$ with $\bm{e}_i$, and most studies reported that $\bm{\omega}$ tends to align with the intermediate eigenvector $\bm{e}_2$, instead of the extensive eigenvector $\bm{e}_1$ \citep{ashurst1987alignment, elsinga2010universal, meneveau2011lagrangian}. Furthermore, when the flow satisfies the homogeneity condition, Betchov established that $\left\langle Q \right\rangle =0$ and $\left\langle R \right\rangle =0$ \citep{betchov1956inequality}. Then from equations \ref{eq:Q_def} and \ref{eq:R_def} we have $\left\langle Q_\omega \right\rangle = -\left\langle Q_s \right\rangle$, and $\left\langle R_s \right\rangle = -\left\langle R_\omega \right\rangle $. It is easy to see that both the enstrophy $\left\langle \omega_i \omega_i \right\rangle$ and the dissipation $\left\langle S_{ij} S_{ij}\right\rangle$ are positive as a result of their quadratic forms. The vortex stretching term $\left\langle \omega_i S_{ij} \omega_j \right\rangle$ is also expected to be positive, as the generation of enstrophy should be positive in turbulent flows. This fact implies that the distribution of vortex stretching should be skewed to the positive part. Moreover, given that $\left\langle R_s \right\rangle = -\left\langle R_\omega \right\rangle = \frac{1}{4} \left\langle \omega_i S_{ij} \omega_j \right\rangle = \left\langle -\lambda_1 \lambda_2 \lambda_3\right\rangle  \geq 0$, we expect the intermediate eigenvalue $\lambda_2$ prefers to be positive statistically \citep{townsend1951fine, betchov1956inequality,meneveau2011lagrangian}.

In the following, we will study the invariants associated with $\mathbf{A}$ as well as $\mathbf{S}$ and $\mathbf{\Omega}$ to explore the impact of polymers on the generation and destruction of small-scale motions in turbulence. We begin by examining their ensemble averages and probability density functions (PDFs) in sections \ref{sec:stat_ensemble} and \ref{sec:stat_pdf}. Next in section \ref{subsec:topo}, we focus on the local flow topology, characterized by the joint PDFs of the aforementioned invariants, and the statistics of eigenvalues and eigenvectors of $\mathbf{S}$ are studied in section \ref{subsec:eigen}. We will find that the statistics are closely related to the vortex sheet structures. Furthermore, our results reveal that the small-scale quantities in VKS are not isotropic, which is discussed in detail in section \ref{sec:stat_anisotropy}. Finally, in section \ref{subsec:vortsheet},  we summarize the key findings of the previous subsections and point out that the observed statistical properties of VGT could be well described by a simple picture: the vortex sheet structures aligned with the axisymmetric axis of the VKS system.

\subsection{Ensemble average of invariants} \label{sec:stat_ensemble}
\begin{figure}
	\centerline{\includegraphics[width=0.8\columnwidth]{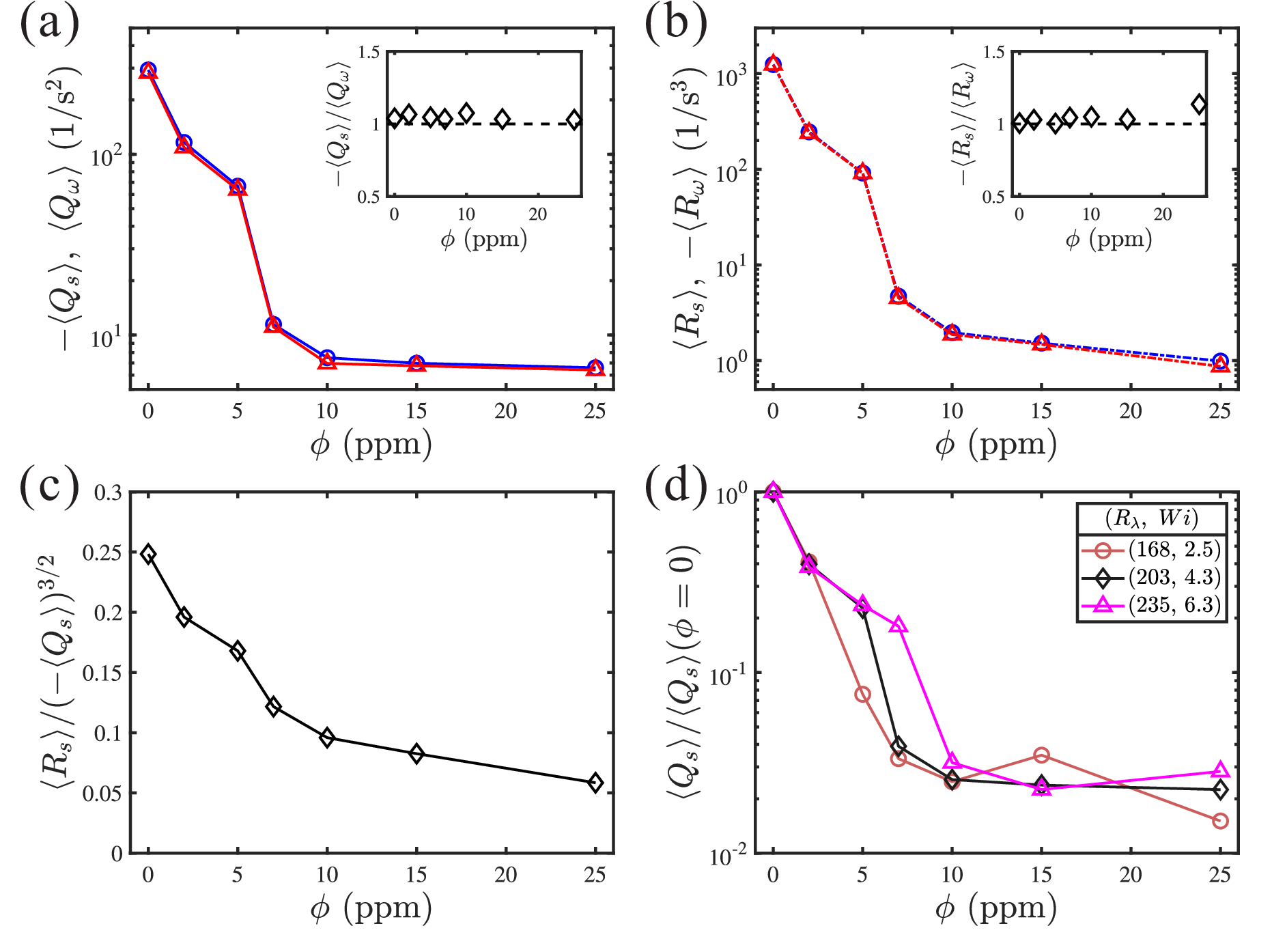}}
	\caption{The ensemble averaged invariants as a function of $\phi$: (a) second-order moments $\left\langle Q_s \right\rangle $ and $\left\langle Q_\omega \right\rangle $, (b) third-order moments $\left\langle R_s \right\rangle $ and $\left\langle R_\omega \right\rangle $. Insets show the ratio between two second- or third-order moments. (c) The skewness of $\mathbf{S}$, $\left\langle R_s \right\rangle/ (-\left\langle Q_s \right\rangle)^{3/2}$. In (a-c) we use dataset $R_\lambda=203$. (d) $\left\langle Q_s \right\rangle $ normalized by the corresponding Newtonian values ($\phi=0$) for three Reynolds numbers $R_\lambda = 168, 203$ and $235$.}
	\label{fig1}
\end{figure}

We first examine the impact of polymers on the ensemble-averaged values of invariants appeared in equations \ref{eq:Q_def} and \ref{eq:R_def}, namely $\left\langle Q_s \right\rangle $, $\left\langle Q_\omega \right\rangle $, $\left\langle R_s \right\rangle $ and $\left\langle R_\omega \right\rangle $. We use the $R_\lambda = 203$ case here. In figure \ref{fig1}(a), we plot the second-order moments $\left\langle Q_s \right\rangle $ and $\left\langle Q_\omega \right\rangle $, and one can see that their magnitudes decrease significantly as the polymer concentration $\phi$ increases, which is consistent with previous studies \citep{de2005homogeneous, liberzon2005turbulence, liberzon2006turbulent, cai2010dns}.  The well-known suppression of the energy cascade in the inertial range due to polymer additions \citep{ouellette2009bulk, xi2013elastic, zhang2021experimental} explains the observed reduction in $-\left\langle Q_s \right\rangle $ and $\left\langle Q_\omega \right\rangle $. Furthermore, given the definition of mean energy dissipation rate $\varepsilon = 2\nu \left\langle S_{ij} S_{ij} \right\rangle $, the results shown in figure \ref{fig1}(a) can be interpreted as drag reduction in bulk turbulence, as reported by previous DNS studies \citep{perlekar2006manifestations, perlekar2010direct, cai2010dns}. In figure \ref{fig1}(b), we plot the ensemble-averaged third-order quantities, strain self-amplification $\left\langle R_s \right\rangle $ and vortex stretching $\left\langle R_\omega \right\rangle $. Similar to the second-order moments, they also decrease significantly with increasing $\phi$. This is consistent with our intuition that the depression of third-order moments leads to the depression of the small-scales in polymeric turbulence \citep{liberzon2005turbulence, liberzon2006turbulent, cocconi2017small}, given that the nonlinear third-order terms contribute to the generation of turbulent system. The insets of \ref{fig1}(a) and (b) show the ratios $-\left\langle Q_s \right\rangle/ \left\langle Q_\omega \right\rangle $ and $-\left\langle R_s \right\rangle / \left\langle R_\omega \right\rangle $, respectively. Remarkably, the data points remain close to 1 within the considered concentration range, which validates the homogeneity condition discussed above \citep{betchov1956inequality}, that is, the fluctuation velocity field could be treated as locally homogeneous, allowing spatial averages to replace ensemble averages. We also notice that some DNS studies introduced the concept of polymer stress and found that while the nonlinear generation terms are suppressed, the polymer stress terms sustain the small scales of turbulence \citep{cai2010dns, watanabe2014power, ur2022effect}, but due to technical limitations, our current experimental setup cannot precisely quantify the impact of polymers on turbulent dynamics.

For the rate-of-strain tensor $\mathbf{S}$, we have discussed its second- and third-order moments,  $\left\langle Q_s \right\rangle $ and $\left\langle R_s \right\rangle $. Both of these moments are significantly suppressed with increasing polymer concentration $\phi$. In addition, figure \ref{fig1}(c) plots the skewness of $\mathbf{S}$, $\left\langle R_s \right\rangle / (-\left\langle Q_s \right\rangle)^{3/2}$, as a function of $\phi$. Within the considered range of $\phi$, the skewness consistently maintains a non-zero value, indicating a skewed distribution of $R_s$ \citep{sreenivasan1997phenomenology, meneveau2011lagrangian} as we mentioned before. However, as $\phi$ increases, the skewness gradually decreases. Consequently, we conclude that polymers not only reduce the magnitude of the second- and third-order moments of $\mathbf{S}$ (or $\mathbf{A}$), but also weaken its skewed distribution, as will be further demonstrated through the PDFs of invariants in the next subsection.

In figure \ref{fig1}(d) we further explore the Reynolds number effect on the moments of VGT in polymeric turbulence, as an example, we plot $\left\langle Q_s \right\rangle $ normalized by their corresponding Newtonian value for various $R_\lambda$ (and also $Wi$). We observed that the normalized $\left\langle Q_s \right\rangle $ decreases differently as a function of polymer concentration $\phi$. Specifically, both curves exhibit a rapid descent at a particular $\phi$ value and as $R_\lambda$ increases, this descent occurs at higher polymer concentration. On the other hand, at sufficiently high $\phi$, the results for all three $R_\lambda$ seem to level off to almost the same extent. Experimental work in VKS with baffled disks \citep{ouellette2009bulk} and smooth disks \citep{zhang2022measured} also reported a similar phenomenon. Recall that a reduction in the amplitude of $\left\langle Q_s \right\rangle $ or $\left\langle Q_\omega \right\rangle $ corresponds to a drag reduction, the results in figure \ref{fig1}(d) as well as previous VKS experiments may represent a maximum drag reduction, suggesting a balanced energy exchange between polymers and turbulent flows.

\subsection{Probability density function of invariants} \label{sec:stat_pdf}

\begin{figure}
	\centerline{\includegraphics[width=0.9\columnwidth]{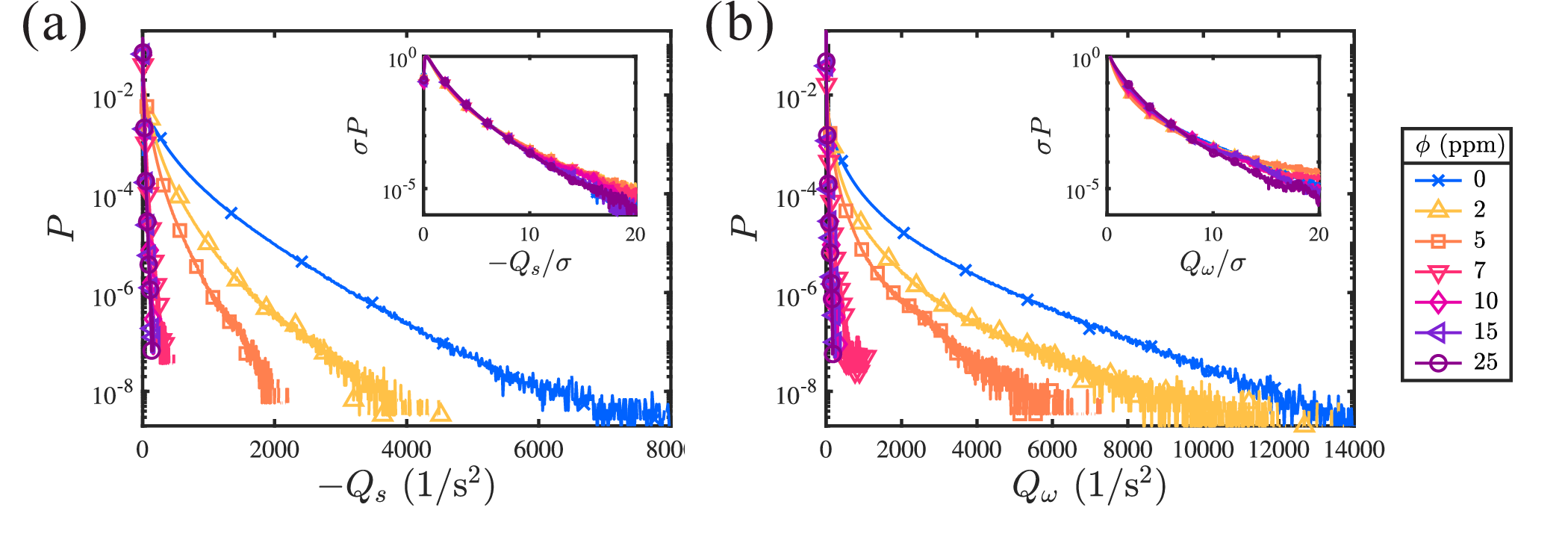}}
	\caption{PDFs of (a) $- Q_s $  and (b) $ Q_\omega $ for different concentrations $\phi$. The insets show the same PDF curves but are normalized by their corresponding standard deviations $\sigma$. Here we use the $R_\lambda=203$ dataset.}
	\label{fig2}
\end{figure}

Compared to the ensemble average, PDF provide more comprehensive information about physical quantities. Figure \ref{fig2} illustrates the PDFs of $- Q_s $ and $ Q_\omega $ at $R_\lambda=203$. Consistent with the reduction of mean value shown in figure \ref{fig1}(a), all the PDF curves in figure \ref{fig2} contract as polymer concentration $\phi$ increases. Intense events with high dissipation or enstrophy, are significantly suppressed, which has been shown in extensive experimental \citep{liberzon2005turbulence,liberzon2006turbulent, zhang2022measured} and numerical \citep{perlekar2010direct, watanabe2013hybrid, cocconi2017small, ur2022effect} research. In addition, we can see that although the mean value of $Q_s$ and $Q_\omega$ balance each other ($- \left\langle Q_s \right\rangle  = \left\langle Q_\omega \right\rangle $), the tails of the PDFs of $Q_\omega$ are always boarder than that of $Q_s$ for different $\phi$, which is consistent with the Newtonian cases \citep{buaria2022vorticity, gotoh2023kinematic}. In the insets of figure \ref{fig2}(a,b) we demonstrate that these PDFs collapse when normalized by their corresponding standard deviations. Our experimental results are consistent with the previous studies \citep{perlekar2010direct, watanabe2013hybrid, zhang2022measured}, revealing a robust functional form of PDF of turbulent energy dissipation in polymeric turbulence with respect to polymer concentration.

\begin{figure}
	\centerline{\includegraphics[width=0.8\columnwidth]{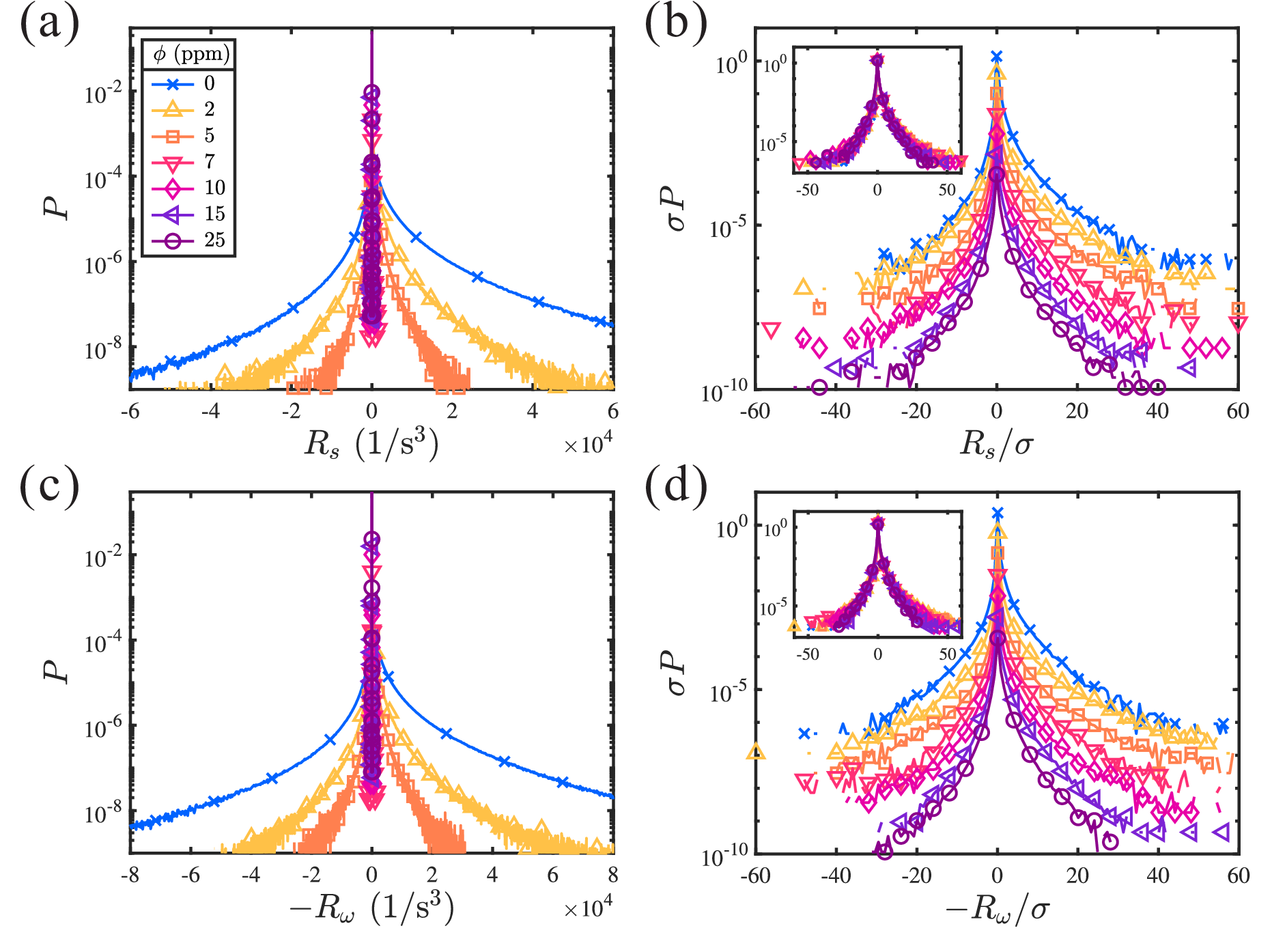}}
	\caption{PDFs of (a) $R_s$ and (c) $-R_\omega$ for different polymer concentrations $\phi$. (b) and (d) present the same data as (a) and (c), but normalized by their corresponding standard deviations, and the curves of higher $\phi$ have been shifted downward by $10^{0.6}$ relative to their lower value neighbors for clarity. The insets show the original, unshifted data. These results are based on the $R_\lambda=203$ dataset.}
	\label{fig3}
\end{figure}

Next, we present the PDFs of third-order invariants $R_s$ and $-R_\omega$ in figure \ref{fig3}. In panels (a) and (c) of figure \ref{fig3}, the PDF curves exhibit a clear trend of shrinking as $\phi$ increases, consistent with the reduction of ensemble average shown in figure \ref{fig1}(b). Similar to the PDFs of $- Q_s $ and $ Q_\omega $ shown in figure \ref{fig2}, the third-order quantities also significantly depressed by the polymer additives, as reported in previous studies \citep{liberzon2005turbulence, liberzon2006turbulent, cai2010dns, cocconi2017small, ur2022effect}. Notably, both $R_s$ and $-R_\omega$ are positively skewed, which remains consistent in polymeric turbulence. And this skewness plays a crucial role in the turbulent energy transfer process. In the context of decaying homogeneous isotropic turbulence (HIT), \cite{cai2010dns} analyzed the changes in production terms and concluded that polymer addition reduces the strength of vortex stretching, leading to the drag-reducing phenomenon.

In figure \ref{fig3}(b,d) we present the same data as in figure \ref{fig3}(a,c) but normalized by their own standard deviations. Note that in the insets, the normalized PDF curves with different $\phi$ exhibit similar form. However, upon closer examination, we observe that right-branch (positive values) curves, representing a positive generation of enstrophy, decrease more rapidly with increasing $\phi$ than the left-branch curves (negative values). In the main figure of figure \ref{fig3}(b,d), these curves for different $\phi$ have been uniformly shifted down $10^{0.6}$ equidistant from each other for clearer comparisons. An interesting finding is that the asymmetric distribution of $R_s$ or $-R_\omega$ is suppressed by polymers, resulting in a decreased value of skewness which is consistent with figure \ref{fig1}(c). \cite{cocconi2017small} first noticed polymers additions have a stronger depressive effect on the positive values of vortex stretching $\omega_i S_{ij} \omega_j$, than on the negative parts. Subsequently, \cite{ur2022effect} studied the orientation between vorticity $\omega_i$ and $S_{ij} \omega_j $. They found that the probability of parallel alignment between these vectors decreases, while the probability of antiparallel increases, which favors negative values of vortex stretching $\omega_i S_{ij} \omega_j$. Our current experiments yield results similar to those shown in figure 9(a) of \cite{ur2022effect} (not shown for simplicity). In summary, polymers suppress intense events of the third-order moments of VGT, with a more substantial impact on positive values of vortex stretching. Our experimental results at higher Reynolds numbers are fully consistent with previous findings at lower Reynolds numbers.

\subsection{Local flow topology} \label{subsec:topo}
\begin{figure}
	\centerline{\includegraphics[width=0.93\columnwidth]{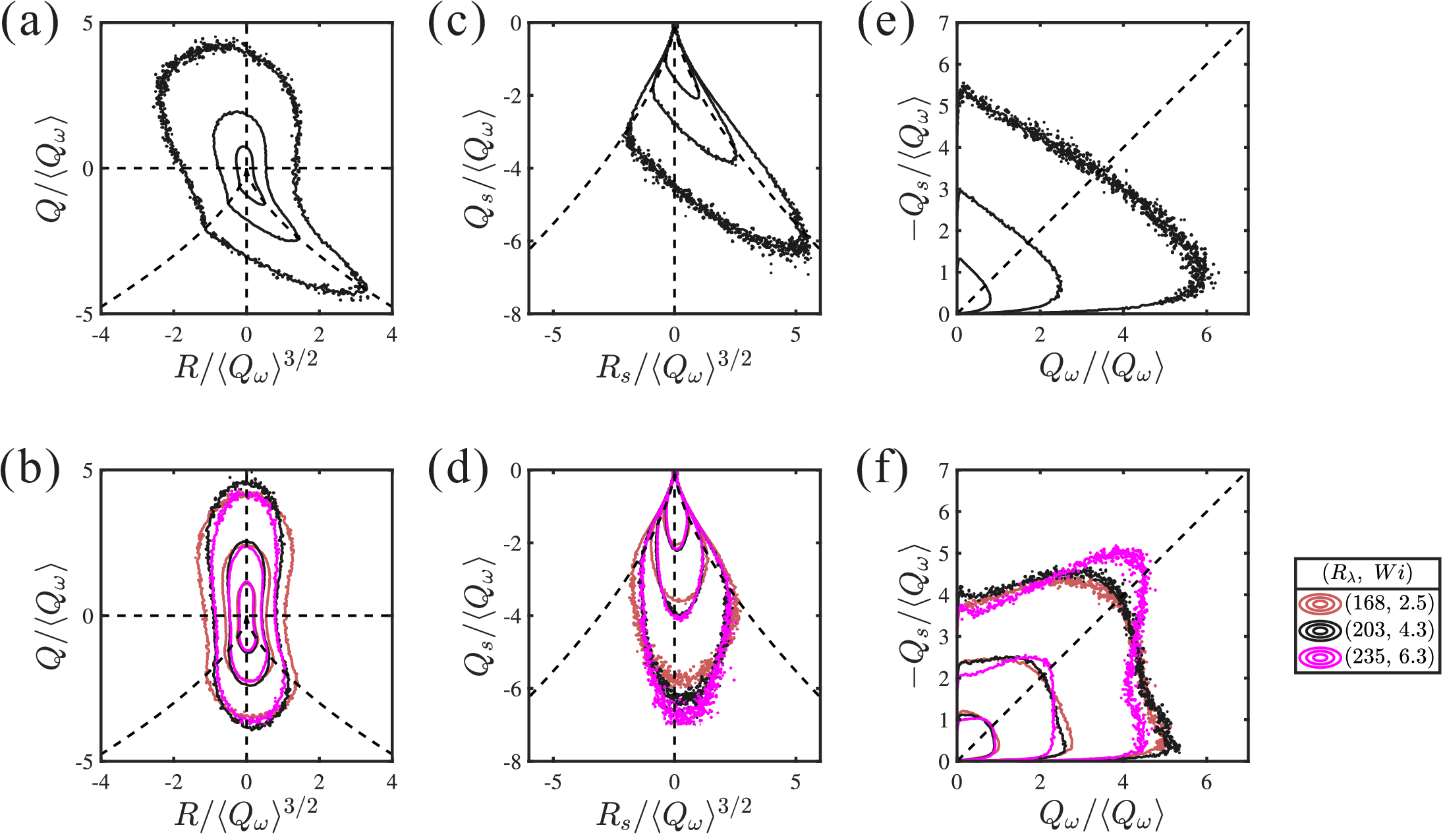}}
	\caption{Joint PDFs of $Q$ and $R$ (a,b), $Q_s$ and $R_s$ (c,d), $Q_s$ and $Q_\omega$ (e,f), nondimentionalized by the second-order moment $\left\langle Q_\omega \right\rangle $. The values of the PDF contours are logarithmic and are $0.01,0.1$ and  $1$ in (a,b), $0.005, 0.05$ and $0.5$ in (c,d) and $0.01,0.1$ and $1$ in (e,f). (a,c,e) correspond to results of the Newtonian case, and (b,d,f) for $\phi=25~\rm ppm$. Black curves correspond to the $R_\lambda=203$ case, and panels (b),(d) and (f) also provide the results from another two Reynolds numbers with $\phi=25~\rm ppm$.}
	\label{fig4}
\end{figure}

In addition to basic statistics such as ensemble averages and PDFs, the impact of polymers on small-scale turbulence can be further investigated through the local flow topology, which could be characterized by the invariants introduced earlier in this section. In figure \ref{fig4}, We present joint PDFs of these invariants, which often exhibit universal features across various turbulent flows. The joint PDF of $Q$ and $R$, $Q_s$ and $R_s$, $Q_s$ and $Q_\omega$ are shown in the left, middle and right columns of the figure, respectively. The upper three panels correspond to the Newtonian case, while the lower three panels represent the polymeric case with a polymer concentration of $\phi=25~\rm ppm$. The black curves correspond to $R_\lambda=203$, and in order to clarify the Reynolds number dependence, we also include the $\phi=25~\rm ppm$  case for two additional Reynolds numbers in the lower three panels.

In figure \ref{fig4}(a,b) we show the joint PDFs between $Q$ and $R$, nondimensionalized by $\left\langle Q_\omega \right\rangle $. Contours with values of  $0.01,0.1$ and $1$ have been selected. The Vieillefosse tails denoted by $D \equiv \frac{27}{4} R^2+Q^3 = 0$ are also shown in the figure, where $D$ represents the discriminant of the characteristic equation of $\mathbf{A}$. The different regions in this $R$-$Q$ map characterize the local flow topology: the $D>0, R<0$ and $D>0, R>0$ regions correspond to local vortex stretching and vortex compression, respectively, while the $D<0, R<0$ and $D<0, R>0$ regions correspond to deformation motion subjected to uniaxial extension and biaxial extension, respectively \citep{vieillefosse1982local,vieillefosse1984internal,cantwell1992exact,meneveau2011lagrangian}. As shown in figure \ref{fig4}(a) for the Newtonian case, one can easily recognize the teardrop shape of PDF, which is consistent with previous experimental results \citep{gomes2014evolution, buxton2017invariants, fiscaletti2022tomographic, warwaruk2024local}. This teardrop shape of PDF on $R$-$Q$ map is believed to be a universal phenomenon for small scales in turbulent flows \citep{elsinga2010universal, meneveau2011lagrangian}.

However, after adding polymers, one can clearly see that the contours in figure \ref{fig4}(b) shrink in the $R$ direction, contrasting with the Newtonian case. This attenuation of $R$ indicates the suppression of dissipation and enstrophy generation, consistent with previous discussions on the depression of mean values and PDFs of $R_s$ and $R_\omega$. Interestingly, our findings here are significantly different from previous results reported by \cite{liberzon2006turbulent} and \cite{perlekar2010direct}. They observed that the teardrop shape still exists on the $R$-$Q$ map for polymeric turbulence,  albeit reduced in area. This dramatic change of $R$-$Q$ PDFs might be understood by the polymer dynamics in turbulent flow. It is expected that the coiled polymer molecule can be stretched by the flow into long chain structures \citep{white2008mechanics}, as clearly shown by a number of numerical simulations \citep{terrapon2004simulated, peters2007two, watanabe2010coil}. In \cite{watanabe2010coil}, by plotting the polymer extension conditioned on the different regions of $R$-$Q$ map, they reported that the longest polymer or maximum extension is found preferentially in the upper-left or lower-right regions. These regions correspond to extreme vortex stretching or biaxial extension motion, where polymers experience significant stretching. Analogous to a dumbbell spring, once stretched, the polymer may dampen the flow, especially in these specific regions. We also notice that people have suggested the interaction between polymer and near-wall vortices as an explanation of turbulent drag reduction \citep{dubief2004coherent}, and indeed, people observed depression of VGT, especially $R$ in polymeric wall turbulence \citep{mortimer2022prediction, warwaruk2024local}, which might be a relevant phenomenon for our findings here in bulk turbulence.

In figure \ref{fig4}(c,d), we present the joint PDFs of the nondimensionalized invariants of  $\mathbf{S}$, $Q_s$ and $R_s$, with contours values $0.005, 0.05$ and $0.5$. Since the rate-of-strain tensor $\mathbf{S}$ is real and symmetric, the PDFs should lie below the curve defined by $D_s \equiv \frac{27}{4} R_s^2+Q_s^3$, that is, $D_s$ should be less than $0$, where $D_s$ denotes the discriminant of the characteristic equation of $\mathbf{S}$. In reality, a few events with $D_s>0$ arise due to unavoidable divergence errors in the experiments, which can also be seen in previous experimental results \citep{gomes2014evolution, warwaruk2024local}. These events are rare (approximately one percent of overall single-point statistics) and do not significantly impact our conclusions. In general, this joint PDF of $Q_s$ and $R_s$ characterizes the local topology of deformation motion and given that $R_s = -\lambda_1 \lambda_2 \lambda_3$ in incompressible flows, $R_s$ also marks the sign of $\lambda_2$. In the Newtonian case shown in figure \ref{fig4}(c), the contours are inclined to the right dashed line with $D_s=0, R_s>0$, which corresponds to $\lambda_1 : \lambda_2 : \lambda_3 = 1:1:-2 $. This result is consistent with the previous finding in the literature that the ratios $\lambda_1 : \lambda_2 : \lambda_3 = 3:1:-4 $ or $2:1:-3$ are most probable \citep{ashurst1987alignment, soria1994study, blackburn1996topology}, suggesting a geometry of biaxial stretching. However, in the polymeric case shown in figure \ref{fig4}(d), the contours tend to be closer to the vertical dashed line, corresponding to two-dimensional stretching with $\lambda_1 : \lambda_2 : \lambda_3 = 1:0:-1 $. This is consistent with the results shown in figure \ref{fig3}(b) and \ref{fig4}(b), since the third-order invariant $R_s$ vanishes in this two-dimensional structure.

Next, in figure \ref{fig4}(e,f), we examine the joint PDFs of $Q_s$ and $Q_\omega$, which characterize the flow topology associated with energy dissipation. We have chosen contour values of $0.01$, $0.1$, and $1$. In the Newtonian case shown in figure \ref{fig4}(e), the PDF concentrates toward the horizontal axis. This region represents the event with strong enstrophy but weaker dissipation, corresponding to a vortex tube structure \citep{soria1994study, da2008invariants}. Surprisingly, when polymers are added, the PDF shape changes dramatically, as shown in figure \ref{fig4}(f), where we can see that now the PDF concentrates around the diagonal line. This region represents the events with comparable strength between enstrophy and dissipation, corresponding to vortex sheet structures \citep{soria1994study}.

Combining the results in figure \ref{fig4}, we can conclude that the local flow topology in polymeric turbulence differs significantly from the Newtonian case. The local flow patterns of vortex stretching and biaxial extension, which are abundant in Newtonian fluid turbulence and essential for the stretching of polymer molecules, are greatly suppressed. Additionally, figure \ref{fig4} exhibits statistical characteristics resembling a vortex sheet structure with two-dimensional properties, which has not been reported in other studies. Finally, results with different $R_\lambda$ in figure \ref{fig4}(b,d,f) demonstrate that these findings become more pronounced as $R_\lambda$ increases.

\subsection{Statistics related to eigenvalues and eigenvectors of rate-of-strain tensor} \label{subsec:eigen}

\begin{figure}
	\centerline{\includegraphics[width=0.67\columnwidth]{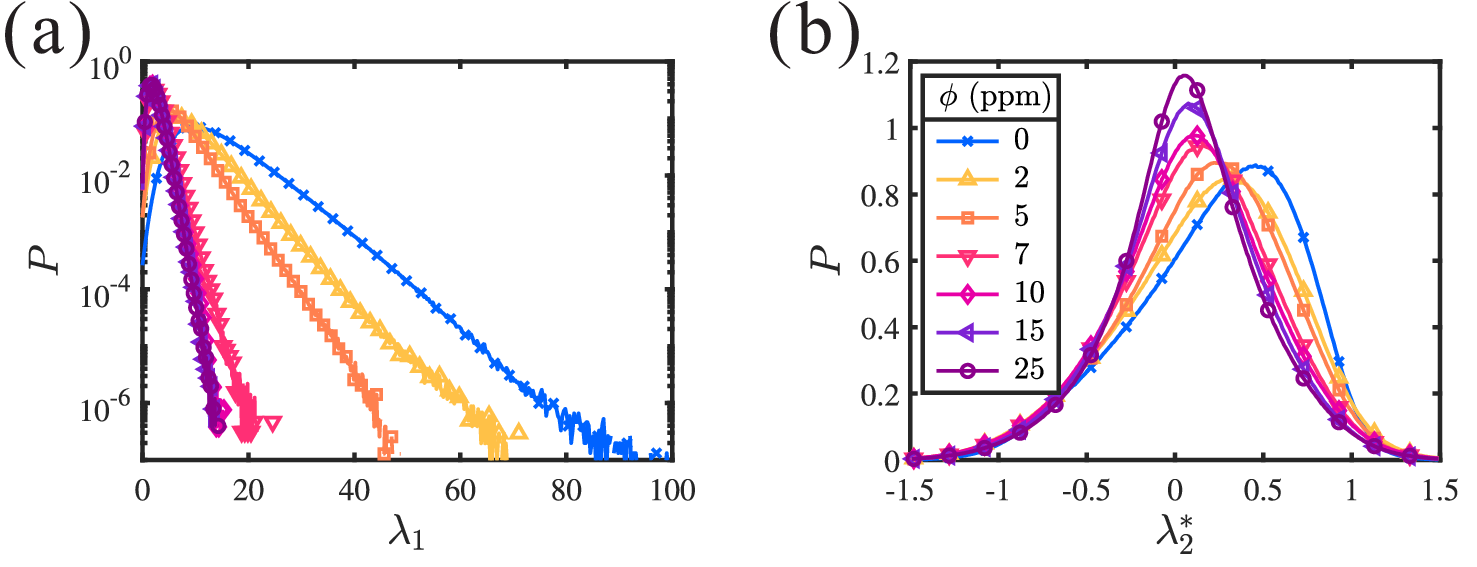}}
	\caption{(a) PDFs of the extensive eigenvalue $\lambda_1$ of $\mathbf{S}$ for different values of $\phi$. (b) PDFs of the normalized intermediate eigenvalue $\lambda_2^\ast=\sqrt{6}\lambda_2/\sqrt{\lambda_1^2+\lambda_2^2+\lambda_3^2}$ for different values of $\phi$. Notice that our measured VGTs do not fully satisfy the divergence-free condition. Thus, the values of $\lambda_2^\ast$ do not completely fall in the range $[-1,1]$ satisfied by incompressible flows. Dataset $R_\lambda=203$ is used here.}
	\label{fig5}
\end{figure}

The joint PDFs of $Q_s$ and $R_s$ shown in figure \ref{fig4}(c,d) suggest that the two-dimensional strain structure, indicated by the ratios $\lambda_1 : \lambda_2 : \lambda_3 = 1: 0 : -1$, dominates the statistics of rate-of-strain tensor $\mathbf{S}$. In this subsection, we further investigate the statistics related to the eigenframe of $\mathbf{S}$ and its dependence on concentration $\phi$. In figure \ref{fig5}, we present the PDFs of $\lambda_1$ and $\lambda_2$ for different values of $\phi$. Notice that the extensive $\lambda_1$ is always positive, and one can see that its PDF shrinks significantly with increasing $\phi$, similar to the trend of invariants shown in section \ref{sec:stat_pdf}. The PDFs of $\lambda_2$ and $\lambda_3$ show similar behaviors, which are not shown here for simplicity. The intermediate $\lambda_2$ could be either positive or negative, and in figure \ref{fig5}(b) we introduce the relative magnitude of $\lambda_2$, defined as $\lambda_2^\ast \equiv \sqrt{6}\lambda_2/\sqrt{\lambda_1^2+\lambda_2^2+\lambda_3^2}$. For the Newtonian case represented by the blue line marked with a cross symbol, we observe an asymmetric distribution with a peak value around $\lambda_2^\ast=0.5$, consistent with previous studies \citep{ashurst1987alignment, lund1994improved, ganapathisubramani2008investigation, buaria2020vortex}. It’s worth noting that our measured VGTs do not fully satisfy the divergence-free condition, which affects the range of values in Figure \ref{fig5}(b). Ideally, incompressible flows should yield values within the range $[-1,1]$. On the other hand, when the polymers are added to the flow, the PDF of $\lambda_2^\ast$ gradually becomes symmetry with respect to the zero point, which is consistent with previous numerical studies on bulk turbulence with polymers \citep{perlekar2010direct, cocconi2017small}. The high probability of $\lambda_2$ near the zero value corresponds to dominance of the eigenvalue distribution  $ \lambda_1 : \lambda_2 :  \lambda_3 = 1:0:-1$, because of $\lambda_1 + \lambda_2 + \lambda_3 = 0$, as we already seen in figure \ref{fig4}(d). Additionally, \cite{liberzon2005turbulence} and \cite{cocconi2017small} listed the ensemble averages of these three eigenvalues in bulk turbulence with and without polymers, which is also consistent with our findings here.

\begin{figure}
	\centerline{\includegraphics[width=.9\columnwidth]{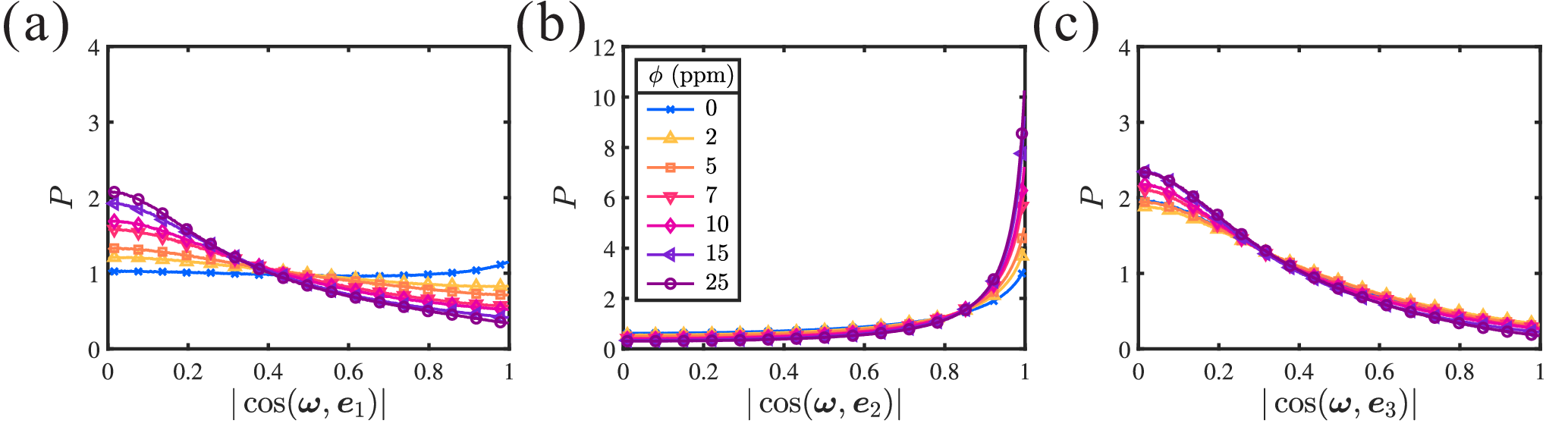}}
	\caption{Alignment of the vorticity vector $\bm{\omega}$ with (a) the first ($\bm{e}_1$), (b) the second ($\bm{e}_2$) and (c) the third ($\bm{e}_3$) eigenvectors of $\mathbf{S}$ for different values of $\phi$. Dataset $R_\lambda=203$ is used here.}
	\label{fig6}
\end{figure}

Apart from the eigenvalues $\lambda_i$ of $\mathbf{S}$, the statistics of their corresponding eigenvectors $\bm{e}_i$ are also of interest, especially their alignments with the vorticity vector $\bm{\omega}$ \citep{elsinga2010universal, meneveau2011lagrangian}. It is well known that in Newtonian turbulence, the vorticity vector has a preferential alignment with the intermediate eigenvector $\bm{e}_2$ \citep{ashurst1987alignment, meneveau2011lagrangian}, which is also observed in our experiments. In figure \ref{fig6}, we plot the PDFs of $\vert\cos(\bm{\omega}, \bm{e}_i)\vert$ for different concentration $\phi$, and in panel (b), the Newtonian result indicated by the blue lines marked with cross symbols, clearly shows a pronounced peak at $\vert\cos(\bm{\omega}, \bm{e}_2)\vert=1$. Additionally, the tendency for vorticity to be randomly aligned with the extensive eigenvector (resulting in a flat PDF for $\vert\cos(\bm{\omega}, \bm{e}_1)\vert$) and perpendicular to the compressive one (with peaks at $\vert\cos(\bm{\omega}, \bm{e}_3)\vert=0$) can also be observed in figure \ref{fig6}(a,c), respectively.

Now we move on to the discussions for polymeric turbulence, and we can see from figure \ref{fig6} that as the concentration $\phi$ increases, the preferential alignment of $\bm{\omega}$ with $\bm{e}_2$ becomes more pronounced, and the peak value of the PDF curve also increases. At the same time, the vorticity vector tends to be orthogonal to $\bm{e}_1$ and $\bm{e}_3$, with a more significant change observed for $\bm{e}_1$. At the highest concentration $\phi=25~\rm ppm$ considered here, the PDFs of $\vert\cos(\bm{\omega}, \bm{e}_1)\vert$ and $\vert\cos(\bm{\omega}, \bm{e}_3)\vert$ are nearly identical, indicating an axisymmetric configuration of vorticity in the eigenframe of strain. Our findings here differ from previous DNS studies, where it was reported that adding polymers has little impact on the alignment between the vorticity and the eigenvectors of the strain tensor $\mathbf{S}$ \citep{watanabe2010coil, cocconi2017small,ur2022effect}. It is possible that the inconsistency between our experimental results and previous DNS studies is related to the large-scale structure. In the current experiments conducted in VKS, the configuration satisfies only the axisymmetry condition rather than isotropy. We will discuss the effect of large-scale anisotropy in detail in the next subsection.

\subsection{Enhanced small-scale anisotropy} \label{sec:stat_anisotropy}

In the preceding subsections, we explored the statistics of invariants related to the VGT and the eigenframe of the $\mathbf{S}$. Now, in this subsection, we move to the discussions on the impact of large-scale effects on VGT statistics, with a particular focus on the influence of large-scale anisotropy. In real fluid systems, such as wall flows and free-shear flows, large-scale motions are inherently anisotropic. Adding polymers to the flow further amplifies this anisotropy. For instance, in boundary layers, the streamwise velocity fluctuation increases with polymer addition, while the transverse fluctuation decreases \citep{warholic1999influence, white2004turbulence}. Therefore, the anisotropy becomes more pronounced compared to the Newtonian counterpart. Similar observations have been made in other flow systems as well \citep{tong1990anisotropy, van1999decay, boffetta2010polymer, lacassagne2019oscillating, peng2023effects, lin2022polymer, wang2025energy, xu2025restoration}. In systems with finite Reynolds numbers, the anisotropy of large-scale motions has a substantial influence on small-scale quantities like VGT, resulting in directional dependence in its statistical properties. The VKS system employed in our study is also anisotropic, satisfying only the axisymmetric condition. As a result, statistical differences arise between the axial and horizontal directions \citep{la2001fluid, voth2002measurement, ouellette2006small}. An intriguing and crucial question then emerges: How does the presence of polymers affect the anisotropy of the VKS flow, particularly the properties of VGT we investigate in this work?

\begin{figure}
	\centerline{\includegraphics[width=.9\columnwidth]{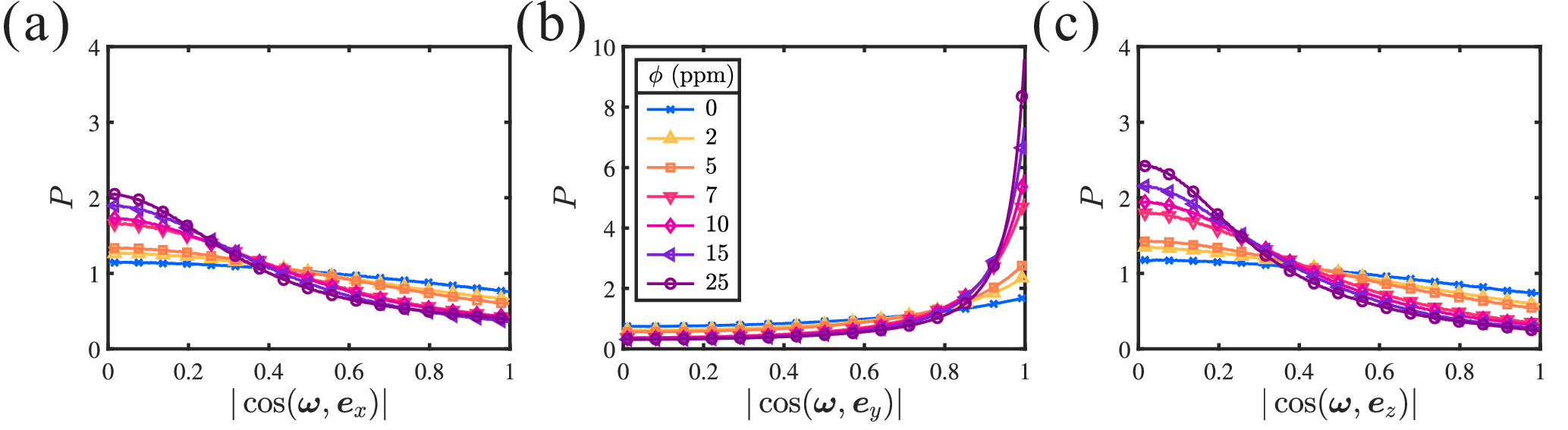}}
	\caption{Alignments of the vorticity vector $\bm{\omega}$ with the laboratory coordinates $\bm{e}_a$ ($a=x,y,z$) for different concentrations $\phi$.  Dataset $R_\lambda=203$ is used here.}
	\label{fig7}
\end{figure}

\begin{figure}
	\centerline{\includegraphics[width=.9\columnwidth]{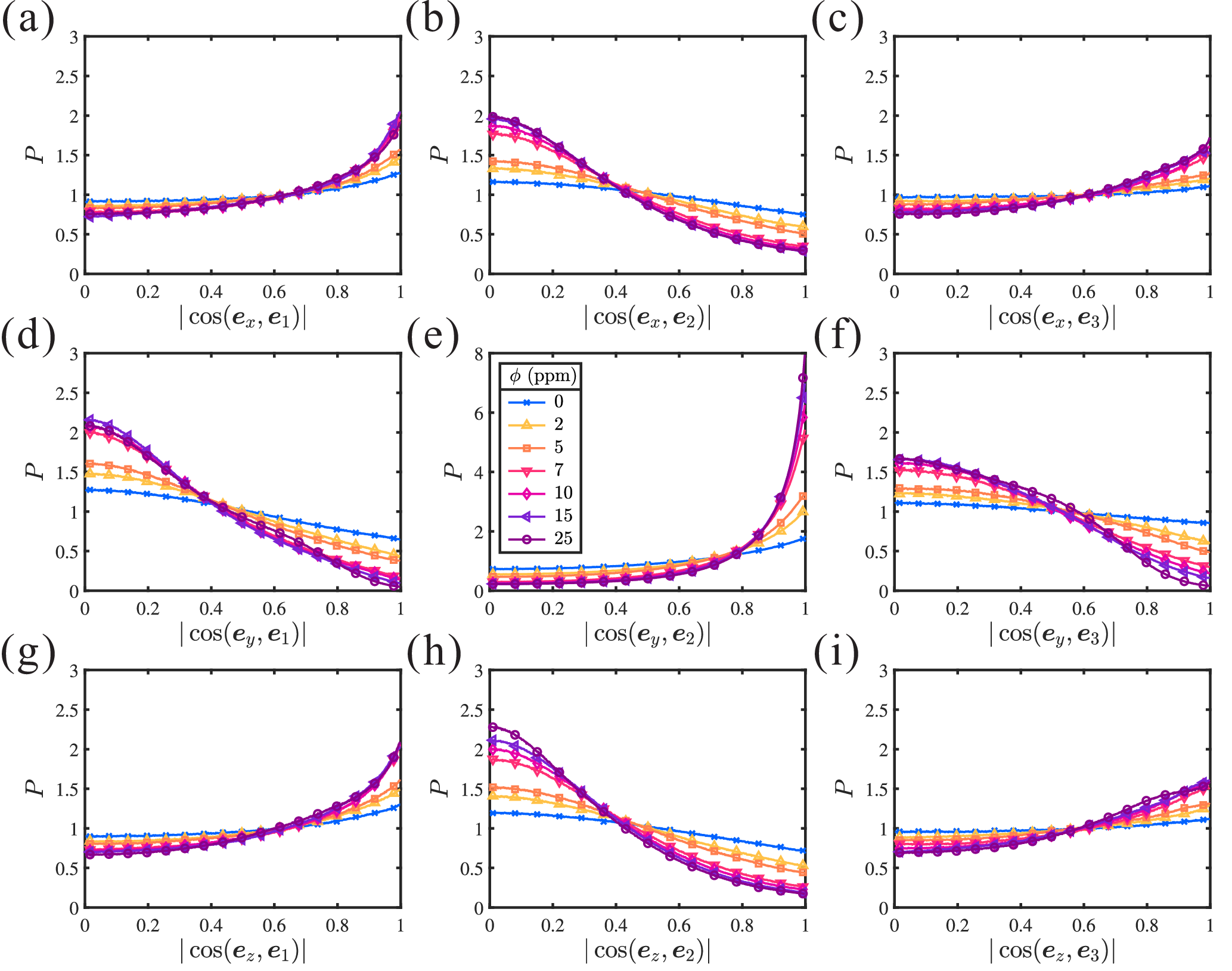}}
	\caption{Alignments of three eigenvectors $\bm{e}_i$ ($i=1,2,3$) of $\mathbf{S}$ with the laboratory coordinates $\bm{e}_a$ ($a=x,y,z$) for different $\phi$. Dataset $R_\lambda=203$ is used here.}
	\label{fig8}
\end{figure}

The anisotropy of the VGT can be characterized by the orientation of the vorticity vector $\bm{\omega}$ and the eigenvectors $\bm{e}_i$ ($i=1,2,3$) of $\mathbf{S}$ in the laboratory frame. We denoted the unit vectors of laboratory coordinates as $\bm{e}_x$, $\bm{e}_y$ and $\bm{e}_z$. In figures \ref{fig7} and \ref{fig8}, we plot the statistics of alignments between vorticity and eigenvectors of strain with respect to the laboratory coordinates for different polymer concentrations. Let’s begin by examining the Newtonian case, represented by the blue lines marked with cross symbols. In an ideal isotropic system, the vectors $\bm{\omega}$ and $\bm{e}_i$ ($i=1,2,3$) should not exhibit any preferential alignment with the laboratory axes $\bm{e}_a$ ($a=x,y,z$). But from figure \ref{fig7} we can see that the vorticity vector $\bm{\omega}$ and the intermediate eigenvector $\bm{e}_2$ tend to align weakly with the vertical direction $\bm{e}_y$, while $\bm{e}_1$ and $\bm{e}_3$ show a weak tendency to be perpendicular to $\bm{e}_y$. We notice that all the results shown in figures \ref{fig7} and \ref{fig8} exhibit symmetry between $\bm{e}_x$ and $\bm{e}_z$, consistent with the axisymmetric nature of the VKS flows. Our findings here agree with the previous experimental measurements \citep{zimmermann2010lagrangian}, revealing subtle yet observable anisotropy in the VGT in Newtonian von K{\'a}rm{\'a}n turbulence.

When polymers are introduced, figures \ref{fig7} and \ref{fig8} show that the trend of alignment gradually strengthens with increasing polymer concentration. Specifically, at the highest concentration $\phi=25~\rm ppm$, $\bm{\omega}$ and $\bm{e}_2$ strongly align with the $\bm{e}_y$, while $\bm{e}_1$ and $\bm{e}_3$ prefer to be perpendicular to $\bm{e}_y$ and lie in the $\bm{e}_x O\bm{e}_z$ plane. Still, the statistics at all values of $\phi$ exhibit symmetry between $\bm{e}_x$ and $\bm{e}_z$. We can clearly see a pronounced enhancement of small-scale anisotropy due to polymer additions. Furthermore, the turbulent energy cascade is known for erasing anisotropy from large scales through inter-scale energy transfer within the inertial range \citep{frisch1995turbulence}, and as the Reynolds number increases, small-scale statistics tend toward isotropy. However, in our study, despite increasing the Reynolds number from $168$ to $235$  (not shown in the figures), we do not observe this trend when polymers are added. This suggests that the enhancement of small-scale anisotropy may be linked to the suppression of energy cascades caused by polymer additions.

In an anisotropic flow with polymer additions, a natural question arises: Does the orientation of polymer molecules have a preferred direction? This issue has been extensively discussed in previous studies \citep{den1997drag, van1999decay, lacassagne2019oscillating, peng2023effects} and confirmed through DNS studies \citep{guimaraes2020direct, mortimer2022prediction}. Even in the case of HIT without a mean flow, observations indicate that polymers tend to be perpendicular to the eigenvector $\bm{e}_3$, and prefer to align with $\bm{e}_2$ at high $Wi$ \citep{watanabe2010coil, valente2014effect, ur2022effect}. Given these findings, it seems plausible that polymers also exhibit a preferred orientation in VKS flows, namely, along the axial direction. However, the mechanism by which oriented polymers enhance existing anisotropy remains a topic for further theoretical exploration, which would inevitably improve our understanding of drag-reducing for wall flows when polymers are present.

\subsection{Vortex sheet structures} \label{subsec:vortsheet}
\begin{figure}
	\centerline{\includegraphics[width=0.35\columnwidth]{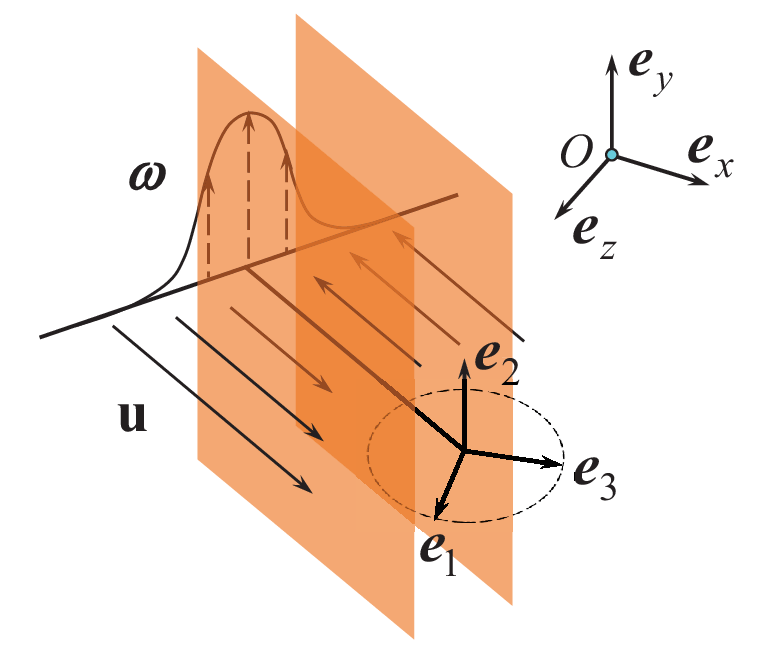}}
	\caption{This sketch illustrates the vortex sheet structure, where $\mathbf{u}$ denotes the velocity field, $\bm{\omega}$ the vorticity, and $\bm{e}_i$ ($i=1,2,3$) the eigenvectors of $\mathbf{S}$. The laboratory coordinates are denoted by orthogonal axes $\bm{e}_a$ ($a=x,y,z$). Notice that the orientation of $\bm{e}_1$ (and thus $ \bm{e}_3$) is random within the  $\bm{e}_x O \bm{e}_z$ plane, as depicted by the dashed circles. This figure is inspired by Fig. 1(b) of \cite{andreotti1997studying}.}
	\label{fig9}
\end{figure}

We have already seen that in a turbulent VKS flow, polymers have strong effects on the statistics of VGT. Recall the results in figures \ref{fig4}-\ref{fig8}; some of them are consistent with previous studies, and some are quite different. To provide a plausible explanation for our findings, let’s summarize the key observations for polymeric VKS turbulence:
\begin {enumerate}
\item Attenuation of the third-order statistics (like $R$ or $R_s$, see figure \ref{fig4}(b,d));
\item The ratios between strain eigenvalues tend to be $ \lambda_1 :  \lambda_2  : \lambda_3 = 1:0:-1$ (see figure \ref{fig4}(d) and \ref{fig5});
\item The coexistence of dissipation $Q_s$ and enstrophy $Q_\omega$ with similar strength (see figure \ref{fig4}(f));
\item The vorticity vector $\bm{\omega}$ aligns preferentially with $\bm{e}_2$ and is perpendicular to both $\bm{e}_1$ and $\bm{e}_3$ (see figure \ref{fig6});
\item The preferential alignment of $\bm{\omega}$ and $\bm{e}_2$ with the aixal direction $\bm{e}_y$, and the symmetry between $\bm{e}_x$ and $\bm{e}_z$ in all statistics (see figures \ref{fig7} and \ref{fig8}).
\end {enumerate}
Surprisingly, we find that a vortex sheet model, like the Burgers vortex layer solution for the Navier–Stokes equations, can perfectly explain the results (i) to (iv) listed above. We sketch Burger's vortex sheet solution (see, for example, Eq. 5 in \cite{andreotti1997studying}) in figure \ref{fig9}, where it becomes evident that Burger's layer corresponds to a simple shear structure. In a simple local shear flow, it is well-established that the magnitudes of the second-order invariants, $Q_s$ and $Q_\omega$, are equal to each other, and the third-order invariants, $R_s$ and $R_\omega$, vanish. Furthermore, the intermediate eigenvector of strain,  $\bm{e}_2$, aligns with the vorticity vector, and the ratios between strain eigenvalues are precisely given by $ \lambda_1 :  \lambda_2  : \lambda_3 = 1:0:-1$. The observation (v) can be attributed to the vortex layer being extended along the $\bm{e}_y$ axis and randomly orientated in the local horizontal plane, as depicted in figure \ref{fig9}. In this case, $\bm{\omega}$ and $\bm{e}_2$ point towards $\bm{e}_y$, while $\bm{e}_1$ and $\bm{e}_3$ randomly orientate in the $\bm{e}_x O \bm{e}_z$ plane.

 In summary, we have discovered that the single-point statistics of VGT in polymeric turbulence can be well described by the local vortex sheet structure, which has not been reported previously. Our experimental results exhibit differences compared with existing literature, where they find that with polymer additions, the shape of $R$-$Q$ PDFs is qualitatively unchanged \citep{liberzon2006turbulent, perlekar2010direct}, and there are no significant differences for vorticity alignment with eigendirections of $\mathbf{S}$ from Newtonian fluids \citep{watanabe2010coil, cocconi2017small, ur2022effect}. This discrepancy may be attributed to the intrinsic anisotropy present in VKS flow, which is absent in the numerical simulations. In this section, we focus solely on single-point statistics. In the subsequent two sections,  by investigating the instantaneous visualization and conditional statistics on vorticity, we will try to identify the flow structures and explore their relation with our proposed model in figure \ref{fig9}.

\section{Structural visualization} \label{sec:visual}
The previous section explored the impact of polymers on the statistics of small-scale turbulence, represented by VGT and related quantities. These statistics are closely related to the local flow topology (see equations \ref{eq:Q_def} and \ref{eq:R_def}), as discussed in section \ref{subsec:topo}. Additionally, figure \ref{fig4} revealed a typical type of flow pattern characterized by the emergence of vortex sheet structures. The aim of the present section is to further investigate the changes in small-scale flow topology after adding polymers by visualizing the flow structures. Specifically, we continue to use $-Q_s$ (representing dissipation) and $Q_\omega$ (representing enstrophy) to reveal the flow structures, which have been extensively adopted in turbulence research \citep{ganapathisubramani2008investigation, gomes2014evolution, buaria2022vorticity}. We present instantaneous results from a selected snapshot. While this selected snapshot contains more representative coherent structures, it is not fundamentally distinct from the other snapshots.

\begin{figure}
	\centerline{\includegraphics[width=.95\columnwidth]{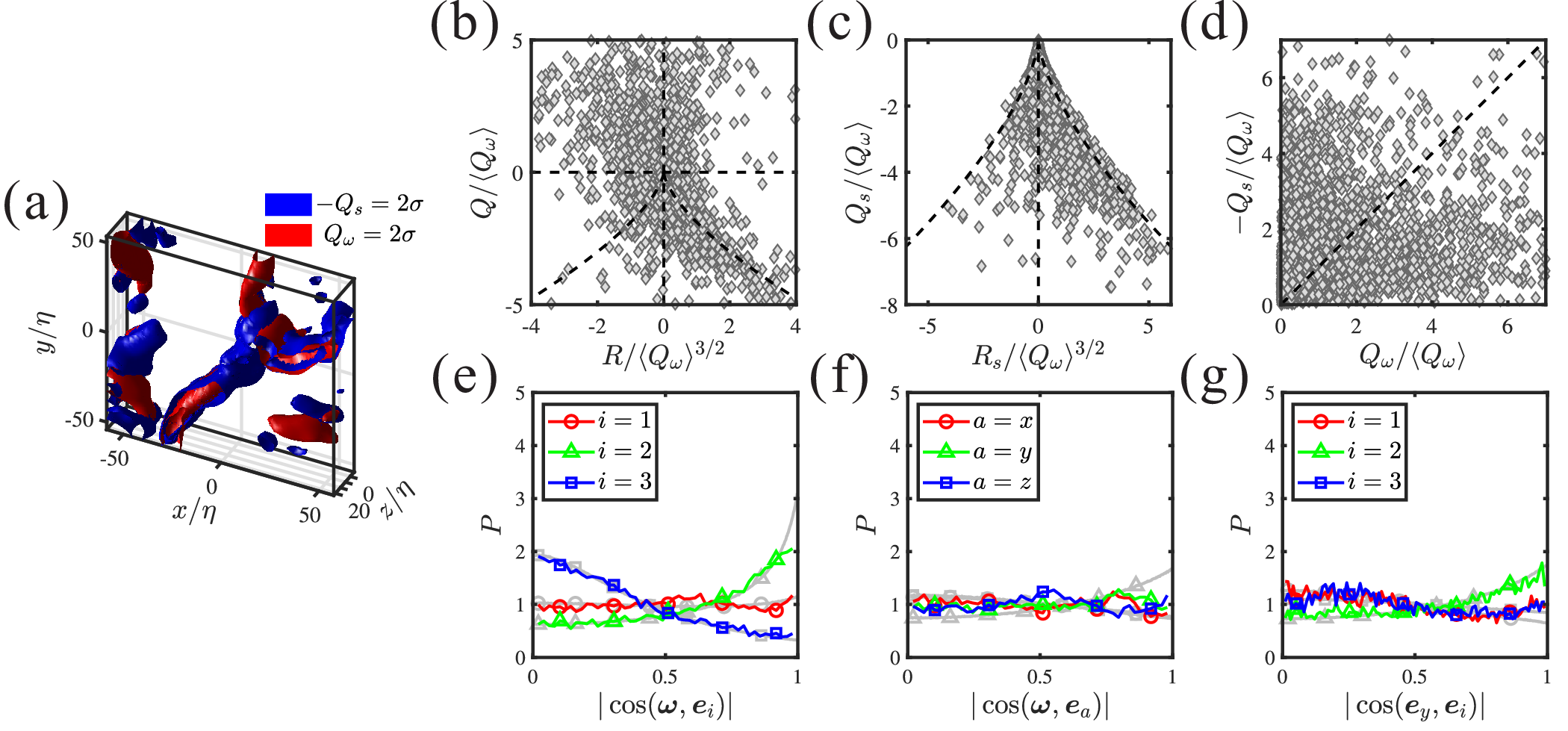}}
	\caption{A representative snapshot and related statistics from the Newtonian case at $R_\lambda=203$. (a) The iso-surfaces of $-Q_s$ (blue) and $Q_\omega$ (red) with a threshold twice their standard deviations. (b)-(d) The scatter plots of $Q - R$, $Q_s - R_s$, and $Q_s - Q_\omega$ corresponding to this snapshot.  Alignments between the vorticity vector and (e) the eigenvectors of $\mathbf{S}$ and (f) the laboratory axis, and (g) the alignments between $\bm{e}_y$ and the eigenvectors of $\mathbf{S}$, all corresponding to the snapshot shown in (a). Notice that the results for full statistics, indicated by the grey curves, are also provided as a reference in (e-g).}
	\label{fig10}
\end{figure}

We begin with the Newtonian case, and the results from a typical snapshot are presented in figure \ref{fig10}. In figure \ref{fig10}(a), we plot the iso-surfaces of extreme values corresponding to twice the standard deviations of $-Q_s$ (blue) and $Q_\omega$ (red). The values of the standard deviations are determined from the full statistics of the Newtonian case at $R_\lambda=203$. We indeed observe the tube-like vortex structures surrounded by the sheet-liked dissipation structures, and these two kinds of structures do not overlap with each other, consistent with previous works \citep{vincent1994dynamics, ganapathisubramani2008investigation, gomes2014evolution, buaria2022vorticity}. We also investigated the effect of the iso-surface threshold in figure \ref{fig10}(a) and found that, within an appropriate range, the shapes of the vortex tube and dissipation structures remain consistent, while their geometric sizes (radius and length) decrease with increasing threshold \citep{buaria2019extreme, buaria2022vorticity, wang2025tomo}.

In figure \ref{fig10}(b-d), with reference to figure \ref{fig4}, we plotted the distributions respectively on the $Q-R$, $Q_s-R_s$ and $Q_s-Q_\omega$ maps corresponding to the snapshot shown in figure \ref{fig10}(a). Given that the statistics contained in figure \ref{fig10}(a) are limited (on the order of $10^4$), we show the scattering plots here instead of the joint PDFs. Nonetheless, by comparing figures \ref{fig10}(b-d) with figure \ref{fig4}, we can see that the characteristic features of topological structures persist in the statistics of a single snapshot since the scattering patterns shown in figure \ref{fig10}(b,c,d) closely resemble those in figure \ref{fig4}(a,c,e), respectively. Consequently, the visualization within a finite-size cube, as depicted in figure \ref{fig10}(a), serves as a good representation of the flow structure in VKS turbulence. This conclusion is further supported by figure \ref{fig10}(e-g), where we plot the alignments between the vorticity vector and the eigenvectors of $\mathbf{S}$, vorticity vector and the laboratory axis, and $\bm{e}_y$ and the eigenvectors of $\mathbf{S}$, all corresponding to the snapshot in \ref{fig10}(a). The full statistics shown in figures \ref{fig6}, \ref{fig7}, and \ref{fig8} are also presented and depicted by grey curves. Notably, the statistics of a single snapshot exhibit similarities to the full statistics, and we could even see a slight preferential alignment of the intermediate eigenvector $\bm{e}_2$ with vertical direction $\bm{e}_y$ for the statistics of a single snapshot in figure \ref{fig10}(g).

\begin{figure}
	\centerline{\includegraphics[width=.95\columnwidth]{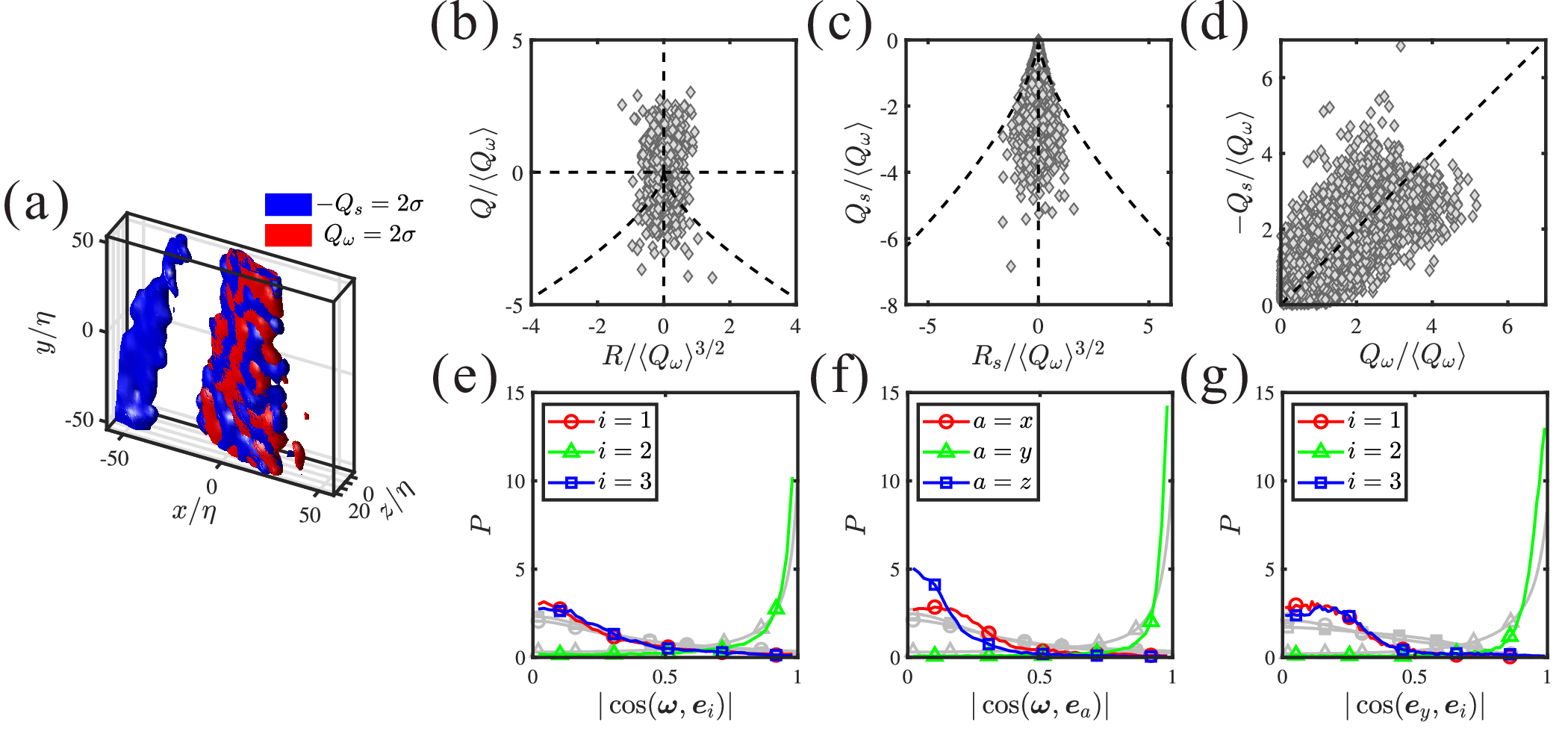}}
	\caption{Quantities same as in figure \ref{fig10}, but for a snapshot chosen from the polymeric case at $\phi=25~\rm ppm$, with $R_\lambda=203$.}
	\label{fig11}
\end{figure}

When adding polymers, a representative snapshot for the $\phi=25~\rm ppm$ case is shown in figure \ref{fig11}, with the same description as in figure \ref{fig10}. Comparing figure \ref{fig11}(a) with figure \ref{fig10}(a), we observe that even a tiny amount of polymer can dramatically alter the distributions of $-Q_s$ and $Q_\omega$, resulting in a sheet-like structure where dissipation and enstrophy coexist in the same region. This is depicted by the intertwined blue and red flow structure in the middle of figure \ref{fig11}(a), which bears some resemblance to the cartoon plotted in figure \ref{fig9}, and supports the conclusion we made from single-point statistics in section \ref{sec:stat}. However, upon examining more snapshots, we found that with the emergence of vortex sheet structure in polymeric turbulence, the vortex tube structure that prevailed in Newtonian turbulence almost disappeared. Notably, although most studies on polymer-laden bulk turbulence reported significant inhibition of the vortex tubes, it is still the dominant structure, and there is no indication for the existence of the vortex sheets \citep{perlekar2006manifestations, cai2010dns, perlekar2010direct, ur2022effect}. An exception is the work by \cite{horiuti2013remarkable}, who observed the emergence of vortex sheet structures upon adding polymers. Furthermore, \cite{horiuti2013remarkable} examined the interaction between polymer stress and the vortex sheets,  noting that the creation of tubes due to the rolling-up of the sheet is attenuated, which leads to the depression of energy cascade. Given that the vortex tube structure is considered as the backbone of turbulence \citep{siggia1981numerical, kerr1985higher, she1990intermittent, douady1991direct, vincent1991spatial, jimenez1993structure}, and the vortex sheets have been associated with the formation of vortex tubes \citep{lundgren1982strained, vincent1994dynamics}, the dominance of vortex sheet over vortex tubes might suggest a suppression effect on small-scale structures of turbulence by adding polymers.

Now, let’s take a closer look at panels (b-g) of figure \ref{fig11} and compare them with the full statistics. From the $R$-$Q$ scattering plot shown in figure \ref{fig11}(b), one can see that the typical teardrop shape disappears, and the distribution of data points becomes almost symmetrical about both the $Q$ and $R$ axes, which resemblance what we see in figure \ref{fig4}(b). Similar trends are observed when comparing figures \ref{fig11}(c) and (d) with figures \ref{fig4}(d) and (f), respectively. The general finding is that compared with the full statistics, the statistics of the single snapshot corresponding to figure \ref{fig11}(a) distribute closer to the theoretical prediction of the vortex sheet model on the invariants plots (see section \ref{subsec:vortsheet}). Recall that the vortex sheet model predicts that the third-order invariants of VGT vanish, the eigenvalues of $\mathbf{S}$ satisfy the ratio $\lambda_1 : \lambda_2 : \lambda_3 = 1:0:-1$ (the middle dashed line in figure \ref{fig11} (c)), and $-Q_s = Q_\omega$ (the dashed line in figure \ref{fig11} (d)). Further confirmation comes from panels (e-g) of figure \ref{fig11}, where the statistics of single snapshots (indicated by colored lines) exhibit more substantial alignment than the full statistics (indicated by grey lines). These observations suggest that the sheet-like structures characterized by high vorticity values play a significant role in the full statistics shown in section \ref{sec:stat}. To verify this conjecture, in the next section, we will explore the statistics of VGT conditioned on different values of vorticity.

\section{Conditional statistics} \label{sec:cond}
In previous sections, we have demonstrated that the small-scale flow topology is significantly altered by the polymer additives. Specifically, the vortex tube structures that prevailed in Newtonian turbulence are replaced by the vortex sheets, which can be seen by the dramatic change of the single-point statistics of VGT and by comparing the visualizations shown in figures \ref{fig10}(a) and \ref{fig11}(a). In figure \ref{fig11}(a), we focus on a snapshot featuring a prominent vortex sheet structure formed by strong vorticity events. Subsequently, the statistics presented in figures \ref{fig11}(b-g) agree better with theoretical predictions from the vortex sheets model than the full statistics (as discussed in the previous section). This fact indicates that the data points close to vortex sheet predictions in the full statistics (see figure \ref{fig4}-\ref{fig8}) are mainly contributed from strong vorticity regions. To provide further insight, here we present the single-point statistics conditioned on different values of vorticity, which complements the visualization of a single snapshot discussed in the last section. We notice that the conditional statistics of VGT and its relation with small-scale flow structure has been extensively studied in the context of Newtonian turbulence in the literature \citep{jimenez1993structure,moisy2004geometry,ganapathisubramani2008investigation,carter2018small,buaria2019extreme,buaria2020vortex,buaria2022vorticity}.

\begin{figure}
	\centerline{\includegraphics[width=.9\columnwidth]{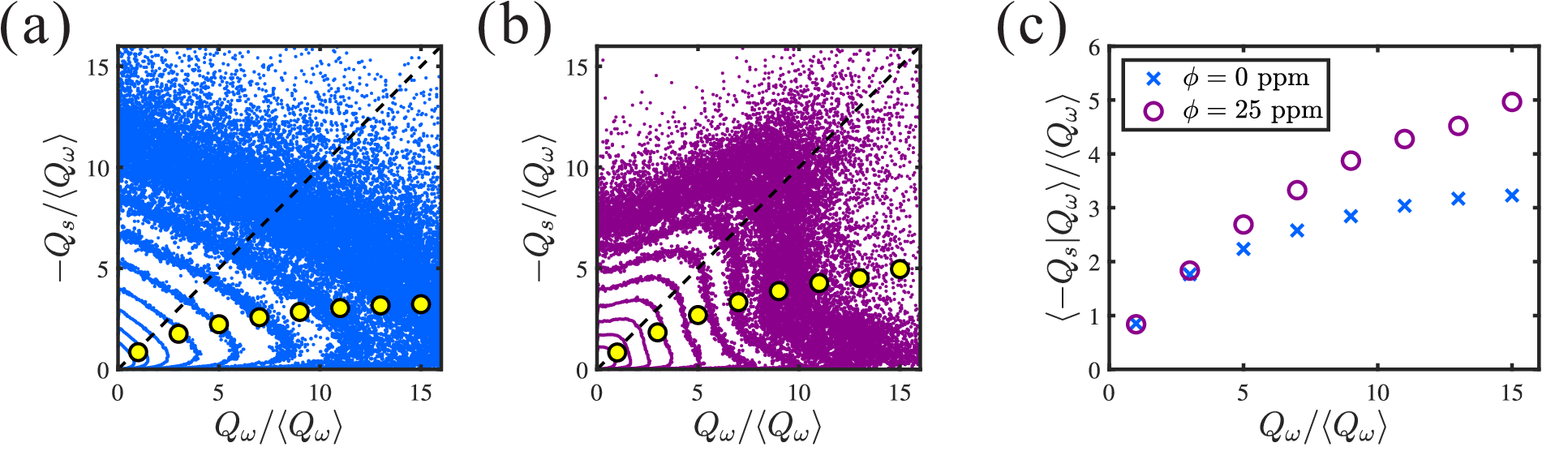}}
	\caption{Joint PDFs of $Q_s$ and $Q_\omega$ for (a) the Newtonian case and (b) the $\phi=25~\rm ppm$ case. Yellow markers denote the averaged value of $-Q_s$ conditioned on the magnitude of $Q_\omega $. The obtained conditional values $\left\langle -Q_s\vert Q_\omega \right\rangle$ from these two cases are summarized in (c). Here we use the datset $R_\lambda=203$.}
	\label{fig12}
\end{figure}

Firstly, in figure \ref{fig12}, we present the joint PDF of $Q_s$ and $Q_\omega$, covering a broader range than that shown in figure \ref{fig4}(e,f). Figure \ref{fig12}(a) represents the Newtonian case; the contours exhibit slight changes as the coordinate values increase, and for the outermost contour, the maximum value of $Q_\omega$ significantly exceeds that of $-Q_s$. We then calculated the mean value of $-Q_s$ conditioned on the strength of $Q_\omega$ within an interval with a width of  $2\left\langle Q_\omega \right\rangle $, as indicated by the yellow dots. The conditional averaged $\left\langle -Q_s\vert Q_\omega \right\rangle$ grows slowly
with increasing $Q_\omega$, and we can see that $Q_s$ is much weaker than $Q_\omega$ in the regions of strong enstrophy, which is due to the strong intermittency of enstrophy \citep{buaria2019extreme,buaria2022vorticity}. As for the $\phi=25~\rm ppm$ case shown in figure \ref{fig12}(b), we observe that the statistics gather along the diagonal line, which suggests the presence of a certain proportion of vortex sheet structures. Similar to figure \ref{fig12}(a), there is an increasing trend in $\left\langle -Q_s\vert Q_\omega \right\rangle$. Furthermore, a direct comparison in figure \ref{fig12}(c) reveals that with polymer additives, $\left\langle -Q_s\vert Q_\omega \right\rangle$ grows faster than the Newtonian case, which is caused by the overwhelming of vortex sheet structures over vortex tubes. Figure \ref{fig12} demonstrates that in regions with higher enstrophy values, we are more likely to encounter the vortex sheet, consistent with our previous findings. Although the vortex sheet structures dominate in polymeric turbulence, it’s important to acknowledge that other structures and even structure-less events also contribute to the full statistics. Nevertheless, this is a crucial foundation for performing conditional statistics, which can offer insights into how polymers impact small-scale turbulence. Next, We will show additional statistical results conditioned on $Q_\omega$.

\begin{figure}
	\centerline{\includegraphics[width=0.67\columnwidth]{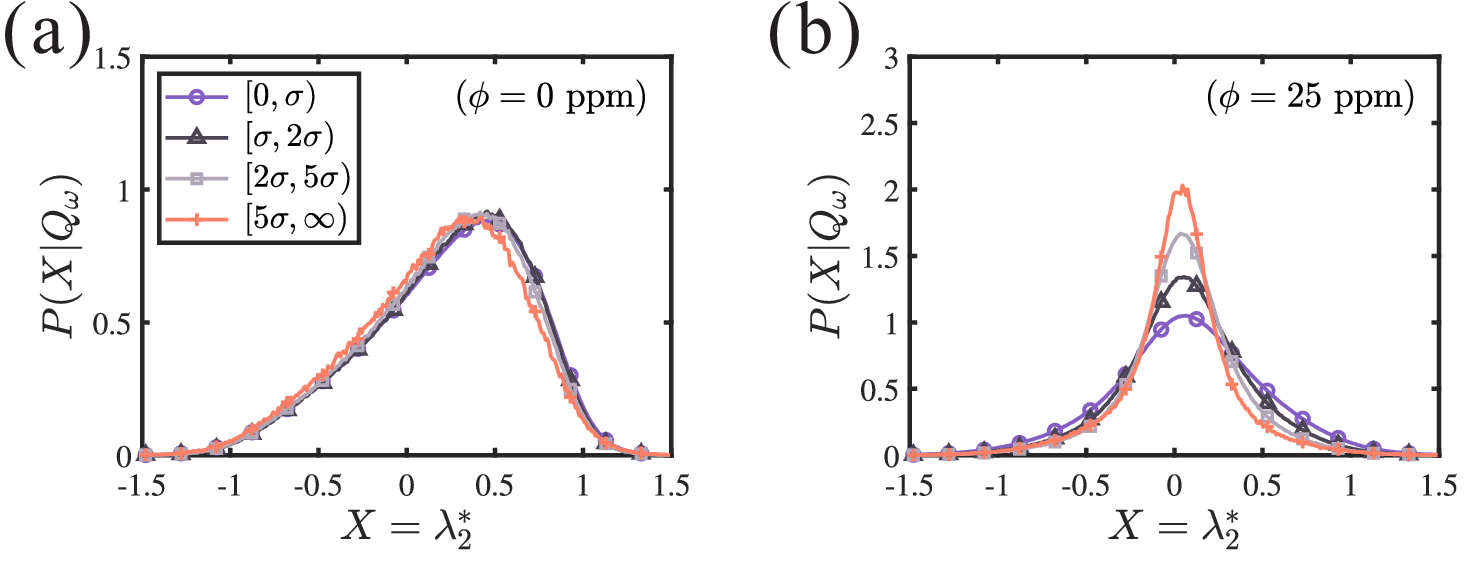}}
	\caption{PDFs of the normalized intermediate eigenvalue $\lambda_2^\ast$ conditioned on the magnitude of $Q_\omega$ for (a) the Newtonian case and (b) the $\phi=25~\rm ppm$ case. The full statistics is divided into four intervals and bounded with values of $0, 1, 2, 5$ times the standard deviation $\sigma$ of $Q_\omega$, as shown in the legend. We use the datset $R_\lambda=203$.}
	\label{fig13}
\end{figure}

In figure \ref{fig13}, referring to the full statistics shown in figure \ref{fig5}(a), we present the PDFs of the normalized intermediate eigenvalue $\lambda_2^\ast$ of the rate-of-strain tensor $\mathbf{S}$, conditioned on different strengths of enstrophy. To achieve this, we divide the entire dataset based on the magnitude of $Q_\omega$, into four intervals: $\left[ 0,\sigma \right) $, $\left[ \sigma,2\sigma \right) $, $\left[ 2\sigma,5\sigma \right) $ and $\left[ 5\sigma,\infty \right) $, where $\sigma$ denotes the standard deviation of $Q_\omega$.
As shown in figure \ref{fig5}(a), for the full statistics, as the concentration $\phi$ increases, the distribution gradually becomes less skewed, which is also observed in other numerical studies \citep{perlekar2010direct, cocconi2017small}. Now considering the conditional statistics, figure \ref{fig13}(a) shows that for Newtonian turbulence, the conditional PDF of $\lambda_2^\ast$ shows no significant variation with respect to vorticity, consistent with the previous experimental \citep{ganapathisubramani2008investigation} and numerical \citep{buaria2020vortex} results. In contrast, the $\phi=25~\rm ppm$ case shown in figure \ref{fig13}(b) exhibit pronounced dependence on vorticity. As the strength of vorticity increases, the PDF of $\lambda_2^\ast$ concentrates more towards zero, indicating a higher probability of the eigenvalue ratio $\lambda_1 : \lambda_2 : \lambda_3 = 1:0:-1 $ for $\mathbf{S}$. Consequently, in regions of intense enstrophy, the local structures of polymeric turbulence are more likely to exhibit two-dimensional features—a phenomenon closely related to the observed vortex sheet structures shown in figure \ref{fig11} (a).

\begin{figure}
	\centerline{\includegraphics[width=.9\columnwidth]{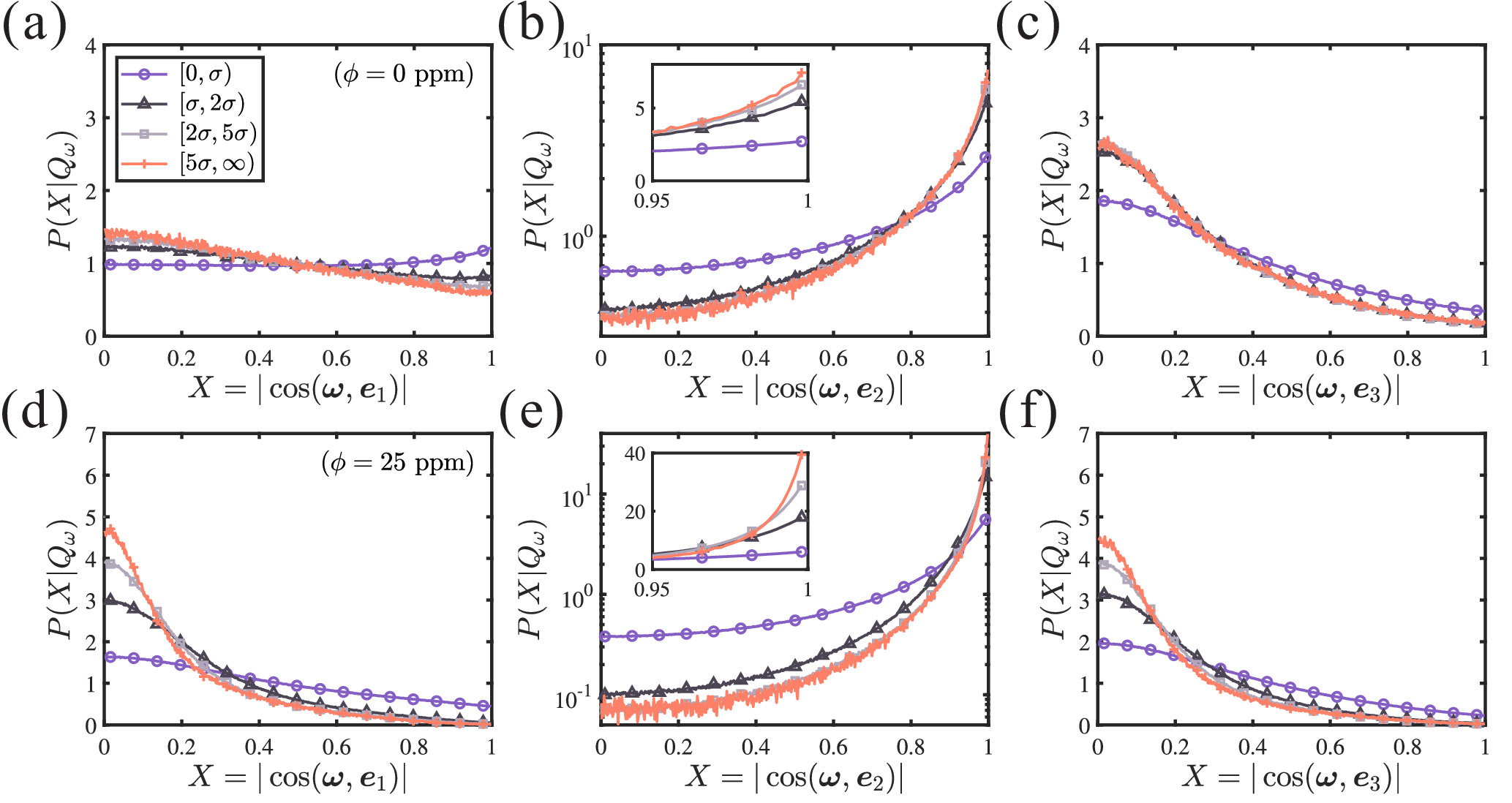}}
	\caption{Alignments of vorticity vector $\bm{\omega}$ with (a,d) the first ($\bm{e}_1$), (b,e) the second ($\bm{e}_2$) and (c,f) the third ($\bm{e}_3$) eigenvector of $\mathbf{S}$ conditioned on the magnitude of $Q_\omega$.  (a-c) are for Newtonian case and (d-f) are for $\phi=25~\rm ppm$. For clarity, the vertical coordinates of (b,e) are logarithmic, and a zoomed-in view is provided in the inset. We use the datset $R_\lambda=203$.}
	\label{fig14}
\end{figure}

 In figure \ref{fig14}, we present conditional statistics regarding the alignment between vorticity and the eigenvectors $\bm{e}_i$ of $\mathbf{S}$ for both the Newtonian and polymeric case. The corresponding full statistics are provided in figure \ref{fig6} earlier. To establish a basic comparison, we begin with the Newtonian case in figure \ref{fig14}(a-c). As the vorticity strength increases, the preferential alignment with $\bm{e}_2$ is intensified, which can be clearly seen in the inset of figure \ref{fig14}(b). The disalignment between vorticity and $\bm{e}_3$ also increases as $Q_\omega$ increases. On the other hand, figure \ref{fig14} (a) shows that with larger values of $Q_\omega$, $\bm{\omega}$ slightly favors being orthogonal to $ \bm{e}_1$. The above observations are consistent with previous DNS results \citep{buaria2020vortex}. In figure \ref{fig6}, we have seen that when adding polymers, $\bm{\omega}$ tends to align with $\bm{e}_2$ and disalign with $\bm{e}_1$ and $\bm{e}_3$. Now from figure \ref{fig14} (d-f) we can clearly see that when $Q_\omega$ increases, this trend of alignment becomes even more pronounced, specifically, the PDF value of $\vert\cos(\bm{\omega}, \bm{e}_2)\vert$ at $\vert\cos(\bm{\omega}, \bm{e}_2)\vert = 1$ increase to a very high value. These results echo those in figure \ref{fig13}, suggesting that regions of intense enstrophy exhibit local flow patterns well-described by the vortex sheet model.

\begin{figure}
	\centerline{\includegraphics[width=.9\columnwidth]{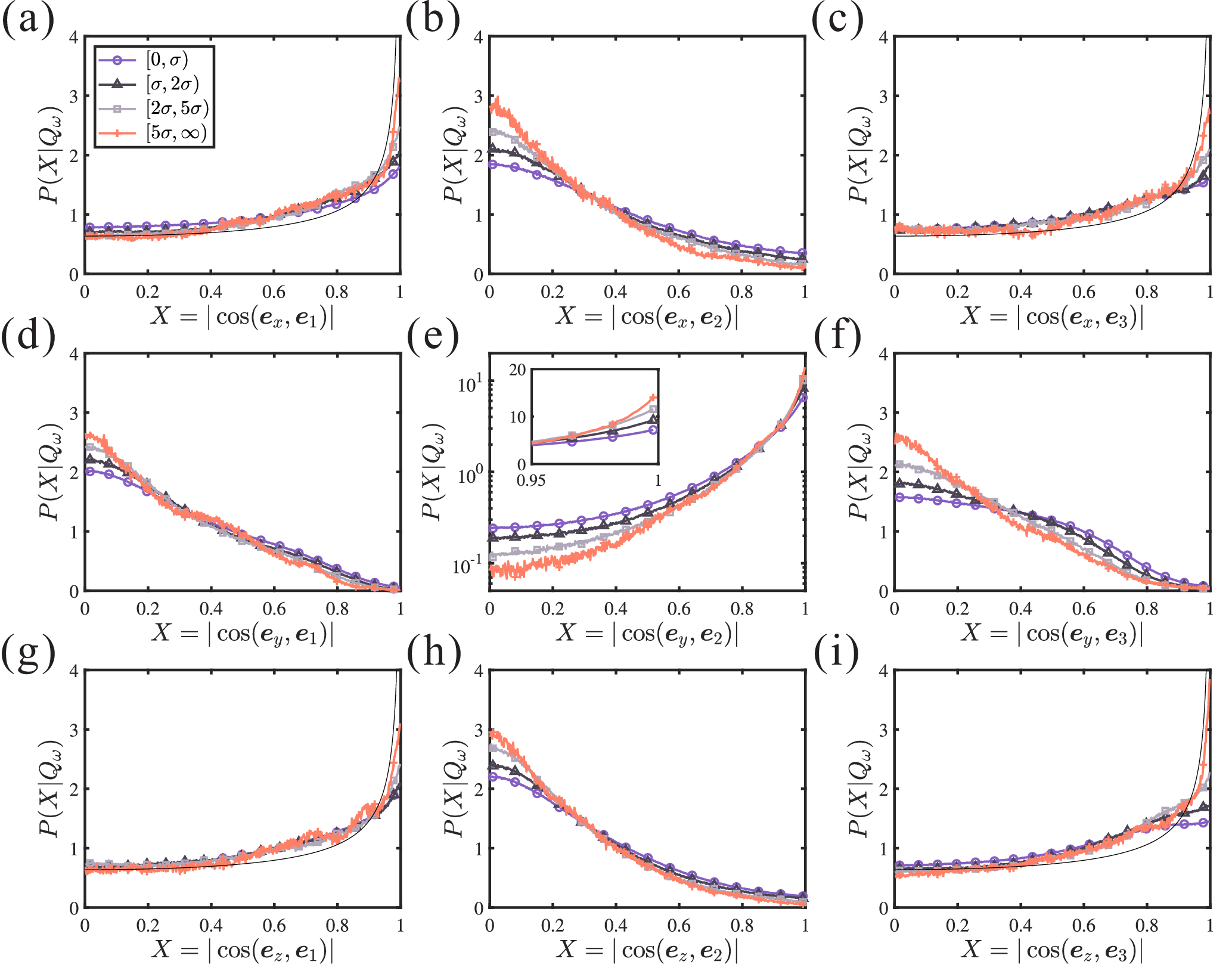}}
	\caption{Alignments of three eigenvectors $\bm{e}_i$ ($i=1,2,3$) of $\mathbf{S}$ with the laboratory coordinates $\bm{e}_a$ ($a=x,y,z$) conditioned on different magnitude of $Q_\omega$. The black curves shown in (a,c,g,i) indicates $ f(X) = \frac{2}{\pi \sqrt{1-X^2}}$ which is obtained from a random orientation between $e_i$ ($i = 1,3$) and $e_a$ ($a = x,z$). For clarity, the vertical coordinate of (e) is logarithmic, and a zoomed-in view is provided in the inset. Data are from the $\phi=25~\rm ppm$ case with $R_\lambda=203$.}
	\label{fig15}
\end{figure}

The enhanced small-scale anisotropy induced by the polymers is further investigated through a conditional analysis of the alignments between eigenvectors of $\mathbf{S}$ with the laboratory coordinates. In VKS flow, where the disks drive the turbulence at the boundary, the small-scale anisotropy is intrinsic. We find that in the Newtonian case, this small-scale anisotropy is nearly unchanged with respect to $Q_\omega$. Figure \ref{fig15} presents the $\phi=25~\rm ppm$ case at $R_\lambda=203$. As the $Q_\omega$ increases, the alignment and disalignment trends become more pronounced and the PDFs for $Q_\omega > 5\sigma$ clearly show stronger alignment compared with the full statistics shown in figure \ref{fig8}. This observation suggests that vortex sheet structures, characterized by intense vorticity regions, tend to align with the axial direction of VKS, resulting in a perfect alignment between $\bm{e}_2$ and the vertical direction $\bm{e}_y$, see figure \ref{fig15} (e). Due to the axisymmetric configuration of the system, the normal vector of the plane formed by the vortex sheet should be randomly oriented in the horizontal $\bm{e}_x O\bm{e}_z$ plane. Consequently, the PDF of the angle between $\bm{e}_i$ ($i = 1,3$) and $\bm{e}_a$ ($a = x,z$) should be the same, as confirmed by figure \ref{fig15} (a,c,g,i). Moreover, this simple picture could even explain the shape of this PDF. In figure \ref{fig15} (a,c,g,i) we plot a black line representing $ f(X) = \frac{2}{\pi \sqrt{1-X^2}}$, which is derived from a uniform distribution $f(\theta) = \frac{1}{2\pi}$, where $\theta$ is the angle between $\bm{e}_i$ ($i = 1,3$) and $\bm{e}_a$ ($a = x,z$), and $X=\vert \cos\theta \vert$. Then, we can see that as $Q_\omega$ increases, the curves of conditional PDFs approach this theoretical prediction, validating our assumptions. Additionally, we calculate the alignments between $\bm{\omega}$ and $\bm{e}_a$ ($a = x,y,z$), and the vorticity exhibits similar behaviors as $\bm{e}_2$, which further supports our observation from the full statistics in figures \ref{fig7} and \ref{fig8}.

Overall, the above results indicate that compared with the full statistics, the statistics of VGT conditioned on strong vorticity tend to concentrate more on the ideal value predicted by a two-dimensional vortex sheet model. Specifically, the conditional average value of $-Q_s$ on $Q_\omega$ is relatively weaker than $Q_\omega$ itself, but enhanced due to the presence of vortex sheet structures.
Further analysis confirms that these structures exhibit nearly two-dimensional flow features and can be effectively described by the idealized vortex sheet model. Additionally, in polymer-laden VKS flow, the vortex sheets exhibit a preferred orientation: they align with the vertical direction while randomly orientating in the horizontal plane, as shown in figure \ref{fig9}.

\section{Concluding remarks} \label{sec:conclusion}
Our study aims to unveil the effects of polymers on small-scale turbulence by conducting Tomo-PIV experiments systematically in polymeric turbulent VKS flow. Previous studies have faced limitations, either due to low Reynolds numbers in numerical studies or incomplete information about VGT in experimental studies. In our current work, based on VKS flow with a relatively high Reynolds number ($R_\lambda=168\text{-}235$), the Tomo-PIV we exploited can provide three-dimensional velocity field with sufficient accuracy to resolve the velocity gradients. This approach allows us to directly analyze the VGT $\mathbf{A}$ and its symmetric and antisymmetric parts. We explore small-scale properties of turbulent flow both with and without polymers, and its dependence on Reynolds number $R_\lambda$ and polymer concentration $\phi$.

We systematically study the statistics of the invariants of VGT, like $Q_s$, $Q_\omega$, $R_s$ and $R_\omega$, etc., which corresponds to dissipation, enstrophy and their generation terms. We observe that the ensemble average of these single-point statistics decreases with increasing concentration $\phi$, as expected. These depression effects are also reflected in their PDFs, in good agreement with most of the experimental or numerical studies that have been reported. Our results also show some similarities between the reduction in $-\left\langle Q_s \right\rangle $ (i.e., drag reduction in bulk turbulence) and drag reduction in polymer-laden wall-bounded flows. 

By investigating small-scale flow topology, we find that the $R$-$Q$ PDF loses its teardrop character in polymeric turbulence, indicating the suppression of vortex stretching and biaxial extension by polymers. The joint PDF between $Q_s$ and $R_s$ is significantly altered, and concentrates around two-dimensional structures with $\lambda_1 : \lambda_2 : \lambda_3 = 1:0:-1 $. The joint PDF of $Q_s$ and $Q_\omega$ show that strong events of dissipation and enstrophy coexist in space. In addition to the statistics of invariants, we also investigate the statistics within the eigenframe of $\mathbf{S}$, especially its alignment with $\bm{\omega}$. The results show that in polymeric turbulence, as the concentration $\phi$ increases,  $\bm{\omega}$ tends to align with the intermediate eigenvector $\bm{e}_2$ of $\mathbf{S}$ and be perpendicular to the other two eigenvectors. These dramatic changes in the statistics of VGT can be well described by the vortex sheet model, indicating the presence of vortex sheet structures in polymer-laden turbulence. These structures replace the vortex tube structures commonly observed in Newtonian turbulence. This clear evidence for the existence of vortex sheet structure has not been reported before. Furthermore, given the anisotropic geometry in the VKS system, the vortex sheet structures have a particular orientation in the flow. To explain our experimental observation, we propose that the typical flow structures in polymeric turbulence are vertically extended vortex sheets randomly oriented in the horizontal plane, as depicted in figure \ref{fig9}. 

The three-dimensional flow fields obtained through Tomo-PIV enable us to visualize structural features. We select two representative flow fields, one from the Newtonian case and another from the polymeric case with concentration $\phi=25~\rm ppm$, and depict the flow structure using iso-surfaces of $Q_s$ and $Q_\omega$. In the Newtonian case, the tube-like vortex structure and the surrounding dissipation regions are clearly visible. However, in the polymer case, the iso-surfaces of two quantities overlap on top of each other and form a sheet-like structure. Moreover, we find that compared with the full statistics in polymeric turbulence, the statistics of the single snapshot with typical vortex sheet structures agree even better with the prediction of the vortex sheet model. This fact indicates the correspondence between intense vorticity structure and single-point statistics. To validate this conjecture, we analyze the conditional statistics of VGT on the enstrophy. It is found that the statistics of VGT, including the statistics of invariants and the alignment between vorticity, eigenvectors of the rate-of-strain tensor, and the laboratory coordinates, agree better with the theoretical prediction of the vortex sheet model at a higher value of enstrophy. This further supports the idea that visually identified vortex sheet structures lead to changes in small-scale topology within polymeric turbulence.

In conclusion, this experimental study enhances our understanding of small-scale turbulence influenced by polymers. In addition to attenuating the magnitude of small-scale statistics, polymers also alter the local flow topology. While Newtonian turbulence predominantly exhibits vortex tube structures at small scales, the introduction of polymers shifts the preference toward vortex sheets. It’s crucial to recognize that the flow we investigated is far away from the wall, which distinguishes it from recent findings in polymer-laden wall-bounded turbulence \citep{mortimer2022prediction, warwaruk2024local}. Although those studies also identified a shear-dominant structure in the viscous wall region, our bulk turbulence with polymers presents a unique context. To fully comprehend the impact of this polymer-induced vortex sheet structure on the turbulent energy cascade and elucidate the emergence of the elastic range, further research is needed.


\backsection[Acknowledgements]{The authors are grateful to Haitao Xu and Alain Pumir for the stimulating discussions.}

\backsection[Funding]{This work is financially supported by the National Natural Science Foundation of China (NSFC) through Grant Nos. 12388101, 12125204, 12202452, and 12402298, and the 111 Project of China through Grant No. B17037.}

\backsection[Declaration of interests]{The authors report no conflict of interest.}


\backsection[Author ORCIDs]{
	
	Feng Wang, https://orcid.org/0000-0001-8466-2754; 
	
	Yi-Bao Zhang, https://orcid.org/0000-0002-4819-0558;
	
	Ping-Fan Yang , https://orcid.org/0000-0003-1949-500X;
	
	Heng-Dong Xi, https://orcid.org/0000-0002-2999-2694.}


\appendix



\bibliographystyle{jfm}
\bibliography{WZYX_ref.bib}


\end{document}